\documentclass[draft]{agujournal2019}
\usepackage{url} 
\usepackage{lineno}
\usepackage[inline]{trackchanges} 
\usepackage{soul}
\draftfalse

\journalname{Radio Science}

\usepackage{amsfonts}
\usepackage{amsmath}
\usepackage{amssymb}
\usepackage{gensymb}
\usepackage{float}
\usepackage[ruled,vlined]{algorithm2e}
\usepackage{aas_macros}
\usepackage{multirow}

\newcommand{\med}[1]{\text{med}\left\{ #1 \right\}}
\newcommand{\MAD}[1]{\text{MAD}\left\{ #1 \right\}}
\newcommand{\mean}[1]{\left\langle #1 \right\rangle}

\begin{document}

%
%

\title{Automated Detection of Antenna Malfunctions in Large-$N$ Interferometers: A Case Study with the Hydrogen Epoch of Reionization Array}

%
%





\authors{
Dara Storer\affil{1},
Joshua S. Dillon\affil{2,\dagger},
Daniel C. Jacobs\affil{3},
Miguel F. Morales\affil{1},
Bryna J. Hazelton\affil{1,4},
Aaron Ewall-Wice\affil{2},
Zara Abdurashidova\affil{2},
James E. Aguirre\affil{6},
Paul Alexander\affil{7},
Zaki S. Ali\affil{2},
Yanga Balfour\affil{8},
Adam P. Beardsley\affil{3,9,\dagger},
Gianni Bernardi\affil{10,11,8},
Tashalee S. Billings\affil{6},
Judd D. Bowman\affil{3},
Richard F. Bradley\affil{12},
Philip Bull\affil{13,14},
Jacob Burba\affil{15},
Steven Carey\affil{7},
Chris L. Carilli\affil{16},
Carina Cheng\affil{2},
David R. DeBoer\affil{17},
Eloy de~Lera~Acedo\affil{7},
Matt Dexter\affil{17},
Scott Dynes\affil{5},
John Ely\affil{7},
Nicolas Fagnoni\affil{7},
Randall Fritz\affil{8},
Steven R. Furlanetto\affil{18},
Kingsley Gale-Sides\affil{7},
Brian Glendenning\affil{16},
Deepthi Gorthi\affil{2},
Bradley Greig\affil{19},
Jasper Grobbelaar\affil{8},
Ziyaad Halday\affil{8},
Jacqueline N. Hewitt\affil{5},
Jack Hickish\affil{17},
Tian Huang\affil{7},
Alec Josaitis\affil{7},
Austin Julius\affil{8},
MacCalvin Kariseb\affil{8},
Nicholas S. Kern\affil{2,5},
Joshua Kerrigan\affil{15},
Piyanat Kittiwisit\affil{14},
Saul A. Kohn\affil{6},
Matthew Kolopanis\affil{3},
Adam Lanman\affil{15},
Paul La~Plante\affil{2,6},
Adrian Liu\affil{21},
Anita Loots\affil{8},
David MacMahon\affil{17},
Lourence Malan\affil{8},
Cresshim Malgas\affil{8},
Zachary E. Martinot\affil{6},
Andrei Mesinger\affil{22},
Mathakane Molewa\affil{8},
Tshegofalang Mosiane\affil{8},
Steven G. Murray\affil{3},
Abraham R. Neben\affil{5},
Bojan Nikolic\affil{7},
Chuneeta Devi Nunhokee\affil{2},
Aaron R. Parsons\affil{2},
Robert Pascua\affil{2,21},
Nipanjana Patra\affil{2},
Samantha Pieterse\affil{8},
Jonathan C. Pober\affil{15},
Nima Razavi-Ghods\affil{7},
Daniel Riley\affil{5},
James Robnett\affil{16},
Kathryn Rosie\affil{8},
Mario G. Santos\affil{8,14},
Peter Sims\affil{21},
Saurabh Singh\affil{21},
Craig Smith\affil{8},
Jianrong Tan\affil{6},
Nithyanandan Thyagarajan\affil{23,\footnotemark{\ddagger},24},
Peter K.~G. Williams\affil{25,26},
Haoxuan Zheng\affil{5}
}

\affiliation{1}{Department of Physics, University of Washington, Seattle, WA}
\affiliation{2}{Department of Astronomy, University of California, Berkeley, CA}
\affiliation{\dagger}{NSF Astronomy and Astrophysics Postdoctoral Fellow}
\affiliation{3}{School of Earth and Space Exploration, Arizona State University, Tempe, AZ}
\affiliation{4}{eScience Institute, University of Washington, Seattle, WA}
\affiliation{5}{Department of Physics, Massachusetts Institute of Technology, Cambridge, MA}
\affiliation{6}{Department of Physics and Astronomy, University of Pennsylvania, Philadelphia, PA}
\affiliation{7}{Cavendish Astrophysics, University of Cambridge, Cambridge, UK}
\affiliation{8}{South African Radio Astronomy Observatory, Black River Park, 2 Fir Street, Observatory, Cape Town, 7925, South Africa}
\affiliation{9}{Department of Physics, Winona State University, Winona, MN}
\affiliation{10}{Department of Physics and Electronics, Rhodes University, PO Box 94, Grahamstown, 6140, South Africa}
\affiliation{11}{INAF-Istituto di Radioastronomia, via Gobetti 101, 40129 Bologna, Italy}
\affiliation{12}{National Radio Astronomy Observatory, Charlottesville, VA}
\affiliation{13}{Queen Mary University London, London E1 4NS, UK}
\affiliation{14}{Department of Physics and Astronomy,  University of Western Cape, Cape Town, 7535, South Africa}
\affiliation{15}{Department of Physics, Brown University, Providence, RI}
\affiliation{16}{National Radio Astronomy Observatory, Socorro, NM}
\affiliation{17}{Radio Astronomy Lab, University of California, Berkeley, CA}
\affiliation{18}{Department of Physics and Astronomy, University of California, Los Angeles, CA}
\affiliation{19}{School of Physics, University of Melbourne, Parkville, VIC 3010, Australia}
\affiliation{21}{Department of Physics and McGill Space Institute, McGill University, 3600 University Street, Montreal, QC H3A 2T8, Canada}
\affiliation{22}{Scuola Normale Superiore, 56126 Pisa, PI, Italy}
\affiliation{23}{National Radio Astronomy Observatory, Socorro, NM 87801, USA}
\affiliation{\footnotemark{\ddagger}}{National Radio Astronomy Observatory Jansky Fellow}
\affiliation{24}{CSIRO, Space and Astronomy, P. O. Box 1130, Bentley, WA 6102, Australia}
\affiliation{25}{Center for Astrophysics, Harvard \& Smithsonian, Cambridge, MA}
\affiliation{26}{American Astronomical Society, Washington, DC}

\correspondingauthor{Dara Storer}{ dstorer@uw.edu}

\begin{abstract}
    We present a framework for identifying and flagging malfunctioning antennas in large radio interferometers. We outline two distinct categories of metrics designed to detect outliers along known failure modes of large arrays: cross-correlation metrics, based on all antenna pairs, and auto-correlation metrics, based solely on individual antennas. We define and motivate the statistical framework for all metrics used, and present tailored visualizations that aid us in clearly identifying new and existing systematics. We implement these techniques using data from 105 antennas in the Hydrogen Epoch of Reionization Array (HERA) as a case study. Finally, we provide a detailed algorithm for implementing these metrics as flagging tools on real data sets.
\end{abstract}

\section{Introduction\label{ch:intro}}

Study of the Epoch of Reionization (EoR) through detection and observation of the 21\,cm emission line from neutral hydrogen will provide critical insights into the formation of the earliest structures of the universe, and help inform understanding of the underlying physics behind galaxy formation and the intergalactic medium  \cite{Furlanetto2006,Morales2010,Pritchard2012}. There are currently several interferometric arrays working to detect the 21\,cm signal, including the the Precision Array for Probing the Epoch of Reionization (PAPER) \cite{PAPER}, the Giant Metrewave Radio Telescope (GMRT; \citeA{GMRT}), the Murchison Widefield Array (MWA; \citeA{MWA}), the LOw Frequency ARray (LOFAR; \citeA{LOFAR}), and the Canadian Hydrogen Intensity Mapping Experiment (CHIME; \citeA{CHIME}), the Hydrogen Epoch of Reionization Array (HERA; \citeA{HERAOverview}), and the Large-Aperture Experiment to Detect the Dark Age (LEDA; \citeA{LEDA}). In addition, there are exciting new experiments on the horizon, including the upcoming Square Kilometer Array (SKA; \citeA{SKA}) and the upcoming Hydrogen Intensity and Real-time Analysis eXperiment (HIRAX; \citeA{HIRAX}). 

The 21\,cm fluctuation signal is very faint; typical models forecast signal amplitudes in the tens of millikelvin, making the signal four to five orders of magnitude fainter than the bright radio foregrounds \cite{Santos2005,Bernardi_2010}. Attempts to measure the power spectrum using radio interferometers must therefore be executed with high sensitivity and precision analysis techniques in order to realistically achieve a detection \cite{Liu_2020}. Achieving sufficient sensitivity requires an interferometer with a large number of antennas observing for months, which introduces a high level of complexity to the system. Therefore, the need for high sensitivity and precision results in thousands of interconnected subsystems that must be commissioned by a relatively small number of people, which poses a significant challenge. Additionally, due to the faintness of the signal, low level systematics that might be deemed negligible in other astronomical applications can have the potential to leak into the power spectrum and obscure the 21\,cm signal. Therefore, systematics must either be resolved, methodically avoided, or directly removed in order to achieve sufficiently clean data. Some examples of contaminants common in these types of interferometers include adverse primary beam effects \cite{Beardsley2016,EW2016,Fagnoni2020,Joseph2019,Chokshi_2021}, internal reflections \cite{EW_2016,Beardsley_2016,Kern2019,Kern2020a,Kern2020b}, radio frequency interference (RFI) \cite{Wilensky2020,Whitler2019}, and any analog or digital systematics resulting from the specific design and configuration of the array and its component electronics \cite{benkevitch2016,LOFAR_2019,Star_2020}. 

In this work we focus on any systematics arising from a malfunction in an individual antenna, component, or subsystem, using data from HERA as a case study to implement and test our methods. While there are some systematics we can avoid using clever analysis techniques (see \citeA{Kern2020a} for example), we manage most systematics by directly removing the affected antennas from the raw data. This requires us to identify and flag any data exhibiting a known malfunction, and develop methodologies for catching new or previously unidentified systematic effects. While the primary goal of flagging data is to produce the cleanest possible data for analysis, it has the added benefit of providing information regarding the scope and character of prevalent issues to the commissioning team, which is essential to our ultimate goal of finding and resolving the source of the problem. The purpose of this work is to outline a framework for identifying and flagging malfunctioning antennas. 

While manual inspection of all data would likely be an effective approach to antenna flagging, for large-$N$ interferometers the data volume poses a problem to this approach. For example, when completed HERA will have 350 individual dishes each with a dual-polarization signal chain including several analog and digital subcomponents. Even just for the 105 elements included in the data used here, manual flagging would involve assessing 22,155 baselines, each of which has 1024 frequency bins and thousands of time integrations. Therefore, the hands-on time involved is neither practical nor reproducible, and so an automated approach is preferred. 

In this paper we present an automated approach to antenna quality assessment and flagging.  Our approach is to design a set of statistical metrics based on common failure modes of the interferometric instruments. We also optimize the metrics to use a limited fraction of the data so they are usable in a real time pipeline. We break these metrics into two categories: cross-correlation metrics (per-antenna values calculated using all baselines), and auto-correlation metrics (per-antenna values calculated using only the auto-correlations). For the duration of this paper we define cross-correlations as correlations between two different antennas, and auto-correlations as the correlation of an antenna with itself. These two methods have complementary advantages. The cross-correlation metrics require a larger data volume, but give us insight into the performance of the whole array and all component subsystems, whereas the auto-correlation metrics are optimized to use a small amount of data, and help assess functionality of individual array components. We outline how each of our metrics is designed to catch one or more known failure modes in the smallest amount of data possible and validate that the automation procedure flags these failures effectively. We also use tools such as simulated noise and comparisons with manual flags to aid in validating our procedure. While these metrics were designed based on HERA data, it is important to note that both the approach and the metrics themselves are applicable to any large interferometric array.

\begin{figure*}[h]
    \centering
    \includegraphics[width=0.99\textwidth]{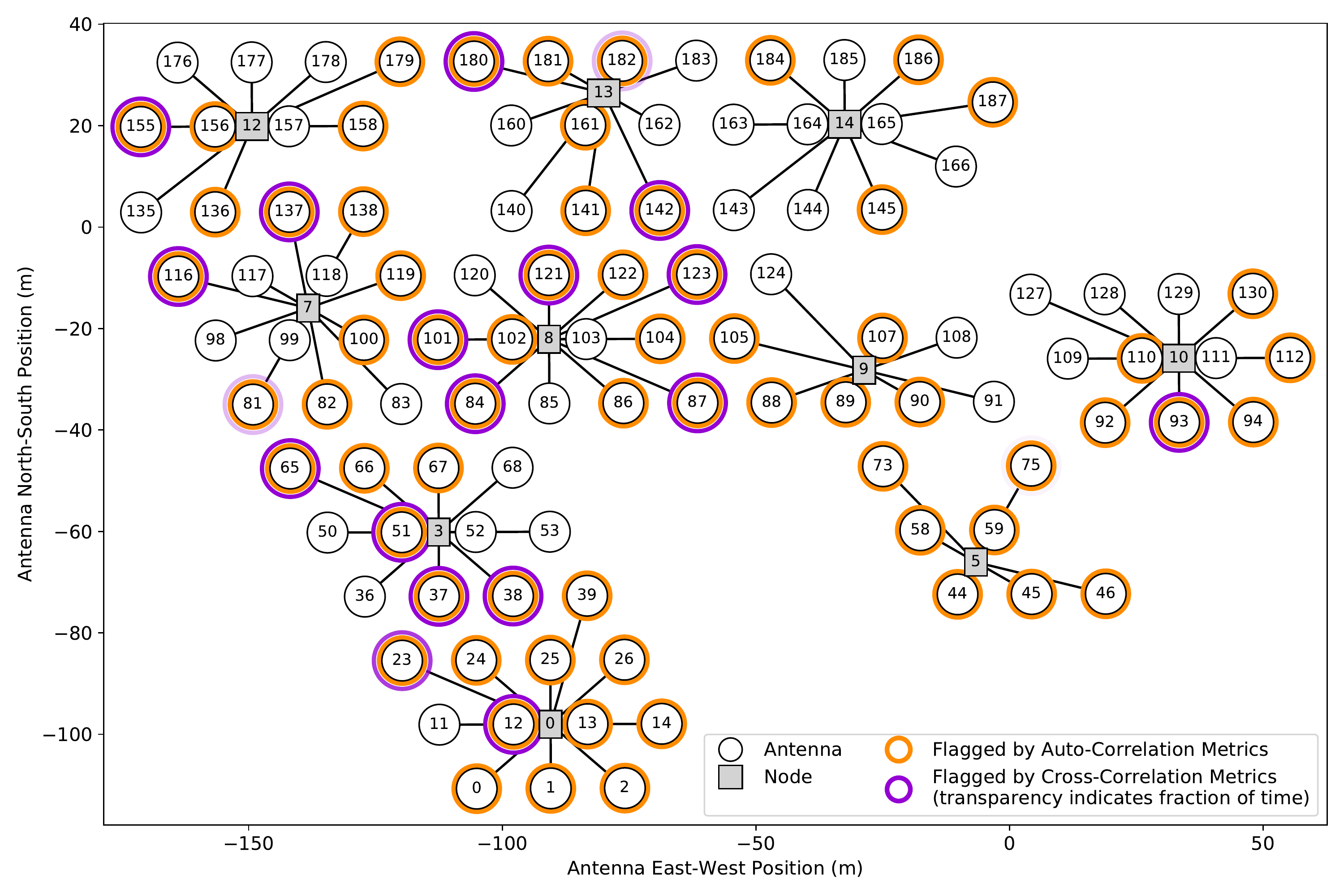}
    \caption{Array layout and antenna quality statuses on Sept 29, 2020 (JD 2459122) as determined by the algorithms laid out in Sections~\ref{ch:ant_metrics} and \ref{ch:auto_metrics}. In HERA, each antenna is connected to a node, which contains amplifiers, digitizers, and the F-engine. Node connections are denoted here by solid black lines. Most of the elements are in the Southwest sector of the split-hexagonal array configuration, with a few in the Northwest and East sectors \cite{dillon_parsons_array_config, HERAOverview}. Only actively instrumented antennas are drawn; many more dishes had been built by this point.}
    \label{fig:array_status}
\end{figure*}

\begin{figure*}[h]
    \centering
    \includegraphics[width=\textwidth]{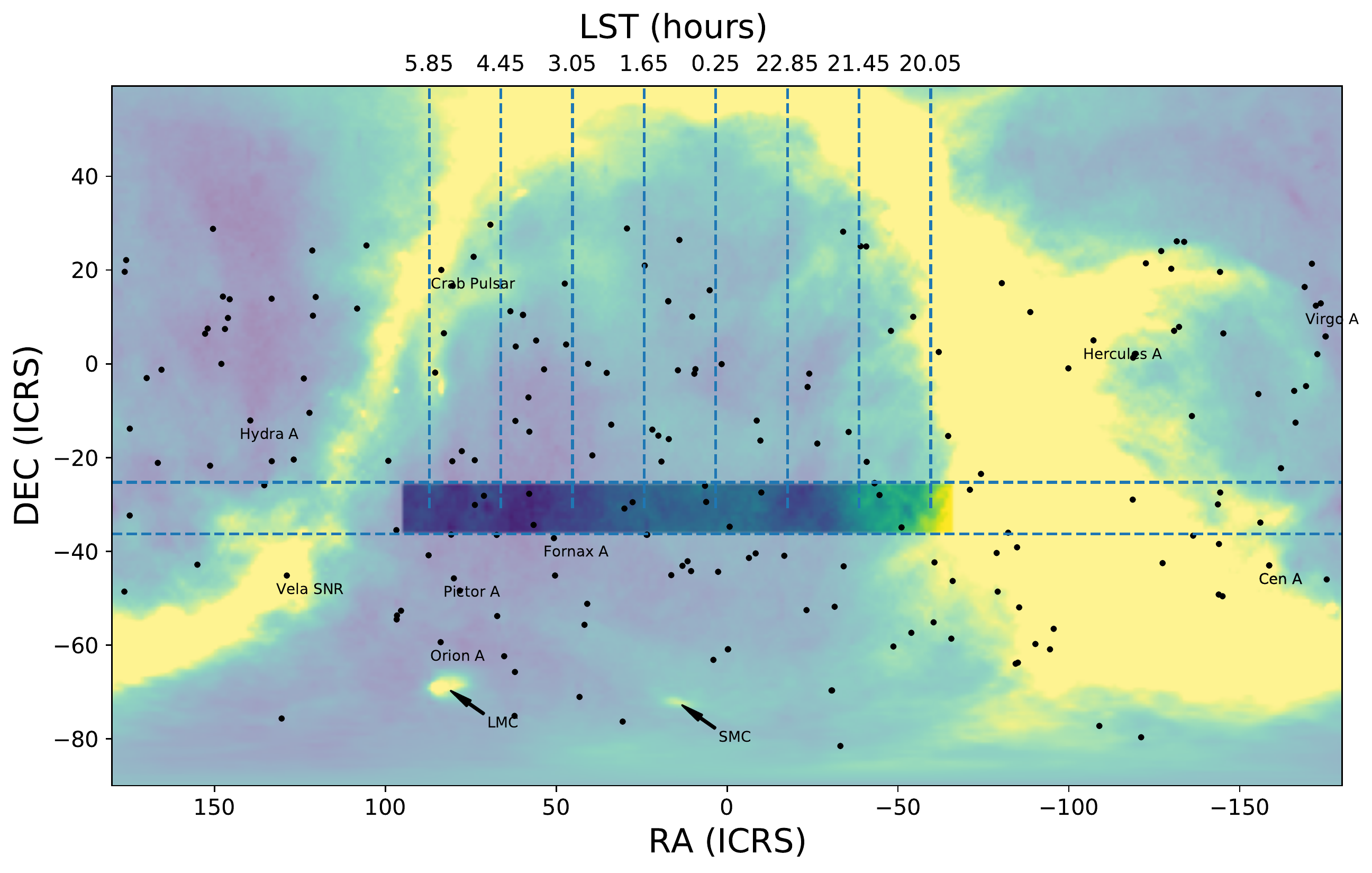}
    \caption{Map of the radio sky \cite{Haslam_2014}, with the HERA observation band for JD 2459122 shaded, based on a Full Width Half Max of 12 degrees. Individual sources shown are those included in the GLEAM 4Jy catalog \cite{GLEAM_4Jy} with a flux greater than 10.9Jy.}
    \label{fig:skymap}
\end{figure*}

The HERA data used in this paper were collected on September 29, 2020 (JD 2459122) when there were 105 antennas online, shown graphically in Figure \ref{fig:array_status}. Note that this data is from the second phase of the HERA array, which uses Vivaldi feeds rather than dipoles, along with other changes, and differs significantly from the phase one data analyzed in \citeA{H1C_IDR2_Limit}. The HERA receivers are distributed throughout the array in nodes which contain modules for post-amplification, filtering, analog to digital conversion, and a frequency Fourier transform. Each node serves up to 12 antennas. Node clocks are synchronized by a White Rabbit timing network \cite{whiterabbit}. Figure \ref{fig:array_status} illustrates the node architecture overlain with antenna cataloging developed in this paper. These flags were produced using almost ten hours of data from this night. The high fraction of malfunctioning antennas was partly attributable to limited site access due to the COVID-19 pandemic. HERA has no moving parts and performs a drift scan observation of $\sim$10$^\circ$ patch around zenith. The portion of the sky observed on JD 2459122 is shown overlaid on the radio sky in Figure \ref{fig:skymap}. 

This paper is organized as follows. In Section \ref{ch:ant_metrics} we outline the two cross-correlation metrics, providing details of their calculation and a demonstration of their utility. We also examine the distribution of the primary cross-correlation metric across the array, and investigate whether systematics are affecting its statistics. In Section \ref{ch:auto_metrics} we introduce four auto-correlation metrics, explaining their necessity, describing their precise statistical formulation, and giving examples of typical and atypical antennas. Finally, in Section \ref{ch:Summary} we summarize our methods and results.

\section{Cross-Correlation Metrics \label{ch:ant_metrics}}

Flagging of misbehaving antennas is necessary in preventing them from impacting calibration, imaging, or power spectrum calculation steps. Here we define a misbehavior to be any feature which makes an antenna unusual when compared to others. In practical terms, the pathologies of antenna malfunction are not limited to the signal chain at the antenna, but could manifest anywhere in the system up to the output of the correlator. Depending on where along the signal chain the pathology lies, we might see evidence of it in the either the auto-correlations, the cross-correlations, or both. For example, if an antenna's timing was out of sync with another's, its auto-correlations might look fine, but its cross-correlations would highlight this systematic. In particular, as an interferometric array grows in size, it is vital to track the health of the entire array, not just the auto-correlations or the cross-correlations in isolation.


In Section \ref{ch:corr_metric} we define a new cross-correlation metric that is aimed at quantifying how well each antenna is correlating with the rest of the array, and we validate this metric with a simulation. Next, in Section \ref{sec:crossPol} we utilize this correlation metric to identify cross-polarized antennas. Finally, in Section \ref{sec:flagging} we outline our specific algorithm for identifying and removing problematic antennas using the cross-correlation metric framework.

\subsection{Identifying Antennas That Do Not Properly Correlate \label{ch:corr_metric}}

Our most generalized metric for assessing antenna function tests how well antennas correlate with each other. There are many reasons antennas might not correlate: one of the gain stages might be broken, cables might be hooked up incorrectly, or not phase-aligned with other functional antennas. Assessment of cross-correlations in uncalibrated data is challenging because the correlations can vary widely depending on the baseline length and sky configuration. In particular, one must be able to tell the difference between baselines that include both the expected sky signal and noise versus baselines that include only noise. A metric which is robust against these and other challenges is the normalized and averaged correlation matrix $C_{ij}$:
\begin{equation}
    C_{ij} \equiv \left\langle \frac{V^\text{even}_{ij} V^\text{odd*}_{ij}}{\left|V^\text{even}_{ij}\right|\left|V^\text{odd}_{ij}\right|} \right\rangle_{t,\nu}
    \label{eq:corr_metric}
\end{equation}
where $\langle \rangle_{t,\nu}$ represents an average over time and frequency, and $V^\text{even}_{ij}$ and $V^\text{odd}_{ij}$ are pairs of measurements of the same sky with independent noise, and $i$ and $j$ are antenna indicies, such that $ij$ represents an individual baseline. This holds for any correlator outputs separated by timescales short enough that the sky will not rotate appreciably, so that we can assume that time adjacent visibilities are observing the same sky signal but with independent noise realizations.\footnote{In HERA's case we are able to utilize our specific correlator output to construct even and odd visibilities that are interleaved on a 100 ms timescale. To explain this, we digress briefly into the output of the HERA correlator. In its last stage of operation, antenna voltage spectra are cross-multiplied and accumulated over 100 ms intervals. These visibilities can be averaged over the full 9.6 second integration before being written to disk. However, in order to improve our estimate of noise and to aid in the estimation of power spectra without a thermal noise bias, we split these 96 spectra into two interleaved groups, even and odd, and sum them independently before writing them to disk. Thus, each is essentially 4.8 seconds of integrated sensitivity, spread over 9.6 seconds of observation. 
}

Division by the visibility amplitude in Equation \ref{eq:corr_metric} minimizes the impact of very bright RFI that might differ between even and odd visibilities and dominate the statistics. We experimented with alternative statistics like a maximum and a median to compress across time and frequency but found that with the normalized correlation a simple average was sufficiently robust.

Due to our chosen normalization, the correlation metric measures the phase correlation between visibilities, and is unaffected by overall amplitudes. If the phases are noise-like, the antennas will be uncorrelated and this value will average down to zero. If $V^\text{even}_{ij}$ and $V^\text{odd}_{ij}$ are strongly correlated, we expect this statistic to be near one. The normalization in Equation~\ref{eq:corr_metric} is particularly useful in mitigating the effects of RFI and imperfect power equalization between antennas. 

\begin{figure*}
    \centering
    \includegraphics[width=0.8\textwidth]{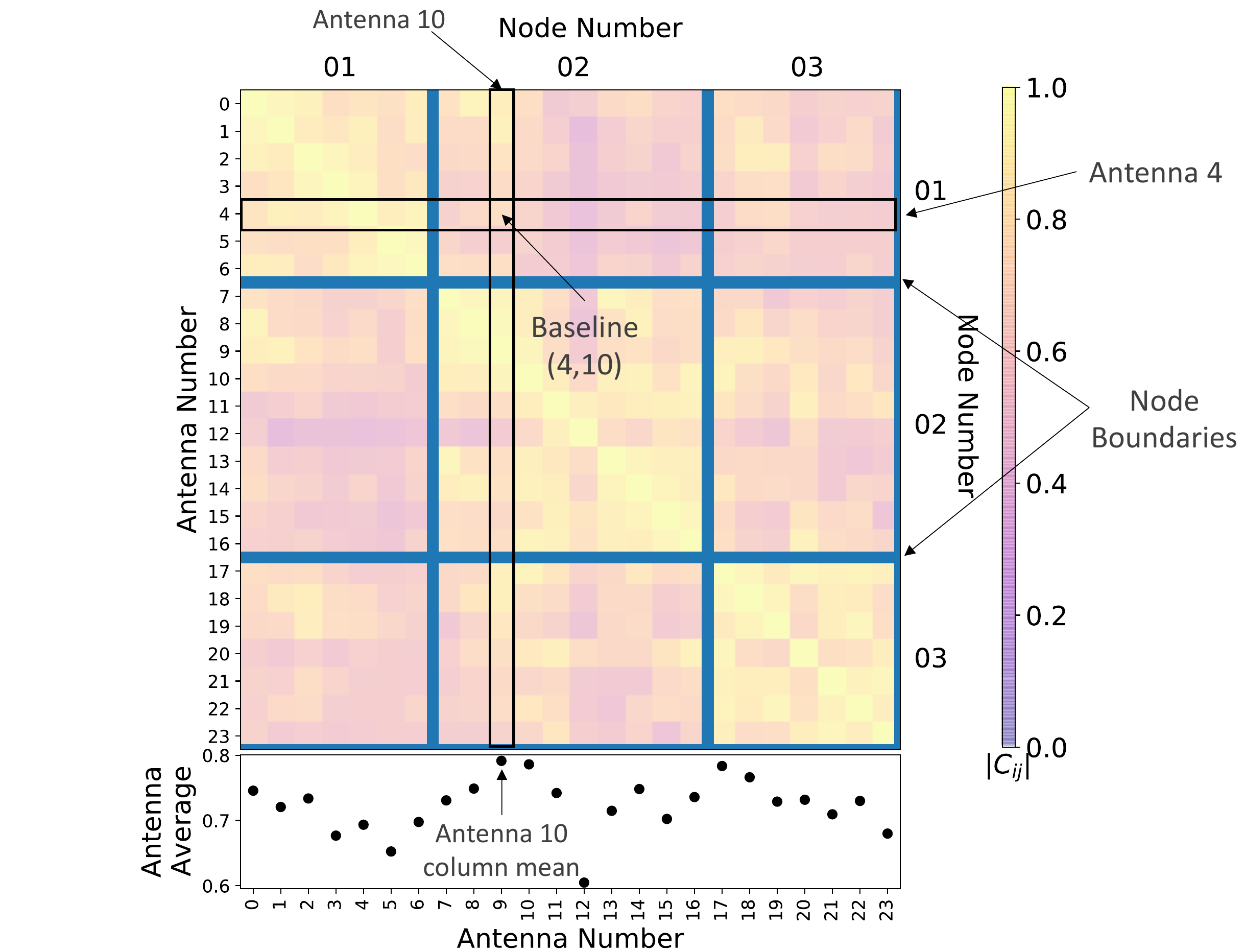}
    \caption{Schematic showing how we visually represent the matrix $C_{ij}$ and the per-antenna metric $C_i$. Each pixel in the matrix represents an individual baseline $ij$, identified by the two antennas that pixel corresponds to. The light blue lines denote the node boundaries, and antennas within each node are additionally sorted by their sub-node level electronic connections. The panel below the matrix shows the per-antenna average, calculated as the column mean for each antenna. (Note that in practice this average is computed iteratively - see Section \ref{sec:flagging}.)}
    \label{fig:sampleMatrix}
\end{figure*}

We can visualize the correlation matrix $C_{ij}$ with each baseline pair $ij$ as an individual pixel, such that the auto-correlations fall along the diagonal. A schematic of this visualization is shown in Figure \ref{fig:sampleMatrix}. To emphasize any patterns related to electronic connectivity, antennas are organized by their node connection, and within that by their sub-node level electronic connections. Node boundaries are denoted by light blue lines. While the nodal structure used here is specific to HERA, the principal of organizing by electronic connectivity is a generalizable technique for highlighting patterns that may be due to systematics in particular parts of the system. Additionally, plotting the matrices in this way allows us to assess the system health on an array-wide level and on an individual antenna level all in one plot, which is increasingly useful as the size of an array grows.

To study the performance of any single antenna it is useful to form a per-antenna cross-correlation metric $C_i$ by averaging over all baselines that include a given antenna:
\begin{equation}
    C_i \equiv \frac{1}{N_\text{ants} - 1} \sum_{j\neq i}{C_{ij}}. \label{eq:C_i}
\end{equation}
where $N_\text{ants}$ is the number of antennas. We calculate this metric separately for all four instrumental visibility polarizations: $NN$, $EE$, $EN$, $NE$. The panels below each matrix in Figure \ref{fig:sampleMatrix} show this per-antenna average correlation metric $C_{ij}$.

\begin{figure*}
    \centering
    \includegraphics[width=1.0\textwidth]{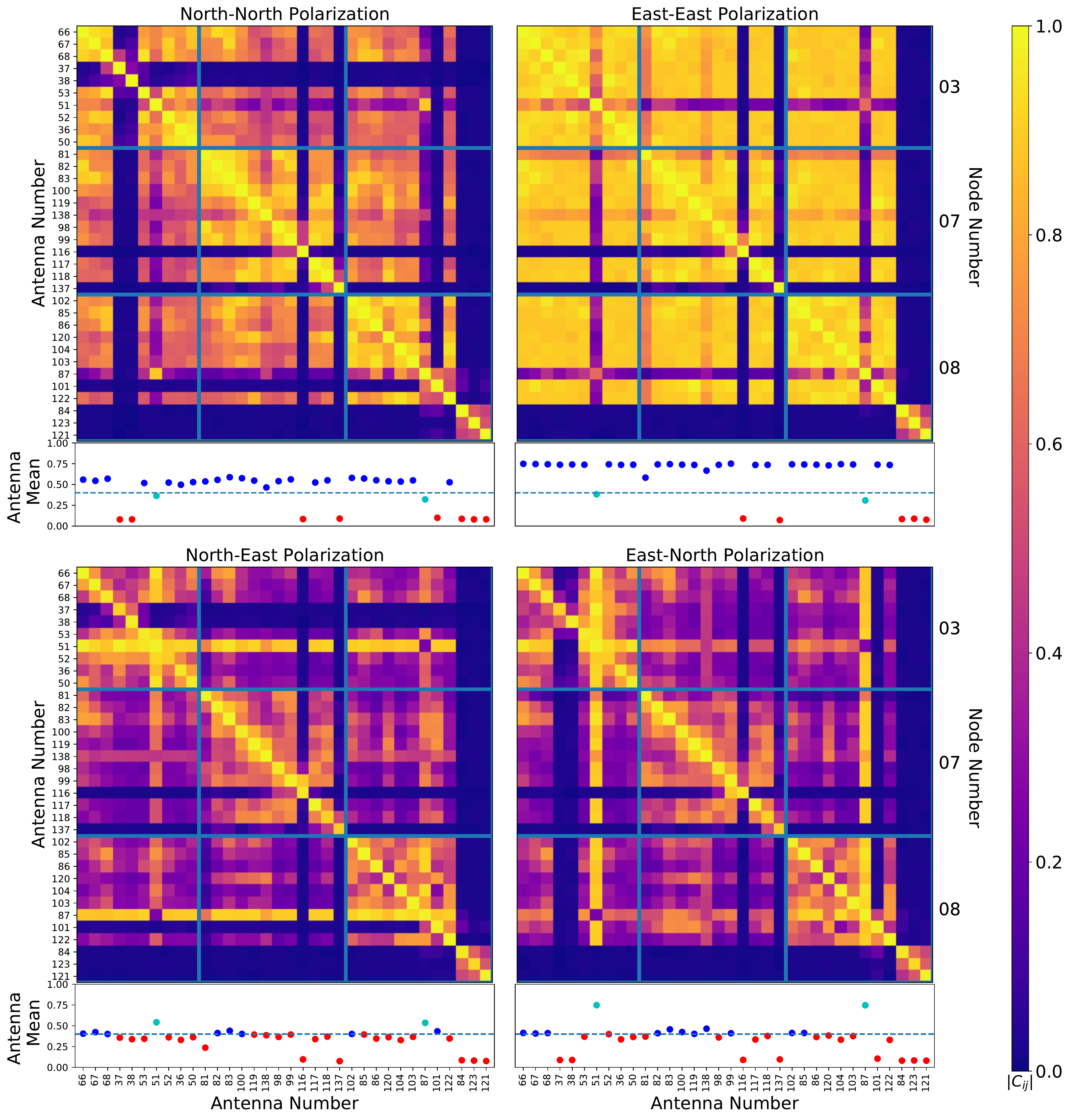}
    \caption{The correlation metric $C_{ij}$ as calculated in Equation \ref{eq:corr_metric}. Light blue lines denote the boundaries between nodes. The per-antenna average metric $C_i$ as calculated in Equation \ref{eq:C_i} is plotted below each matrix. The dashed line indicates the flagging threshold, such that blue dots indicate unflagged antennas, red indicates flagged antennas, and cyan indicates antennas identified as being cross-polarized (see Section \ref{sec:crossPol}).
    Note that we do not use the North-East and East-North polarizations for antenna flagging.}
    \label{fig:corrMatrix}
\end{figure*}

Next, Figure \ref{fig:corrMatrix} shows a visualization of $C_{ij}$ for all four polarizations, using data from a representative subset of the HERA array for simplicity. Here the values have a  bimodal distribution (most obvious in the East-East and North-North polarizations), where most antennas are either showing a consistently low metric value, or are close to the array average. This bimodality is also clear in the lower panels showing the per-antenna metric $C_i$. Here we see more clearly that there is a fairly stable array-level average metric value for each polarization, with a handful of antennas appearing as outliers. The dashed line in the lower panels shows the threshold that is used for antenna flagging, with the points below the threshold marked in red - see Section \ref{sec:flagging} for more on this. There are three primary features to note in Figure \ref{fig:corrMatrix}. First, we see that antennas 51 and 87 are lower than the array average in the North-North and East-East polarizations, but are higher than average in the other two polarizations. Thses points are marked in cyan in the lower panel. The reason for this pathology is that antennas 51 and 87 are cross-polarized, meaning that the cables carrying the East and North polarizations are swapped somewhere along the cable path - this will be discussed further in Section \ref{sec:crossPol}. Second, we can see that the typical value of $C_{ij}$ is higher in the East-East polarization than in the North-North polarization. This is because of the elevated signal-to-noise ratio observed in the East-East polarization due to contributions from the galactic plane and diffuse emission. Lastly, we observe that there appears to be a slight increase in the average metric power for baselines within the same node compared over baselines to antennas in different nodes. We explore this effect in the next section.

\begin{figure}
    \centering
    \includegraphics[width=0.45\textwidth]{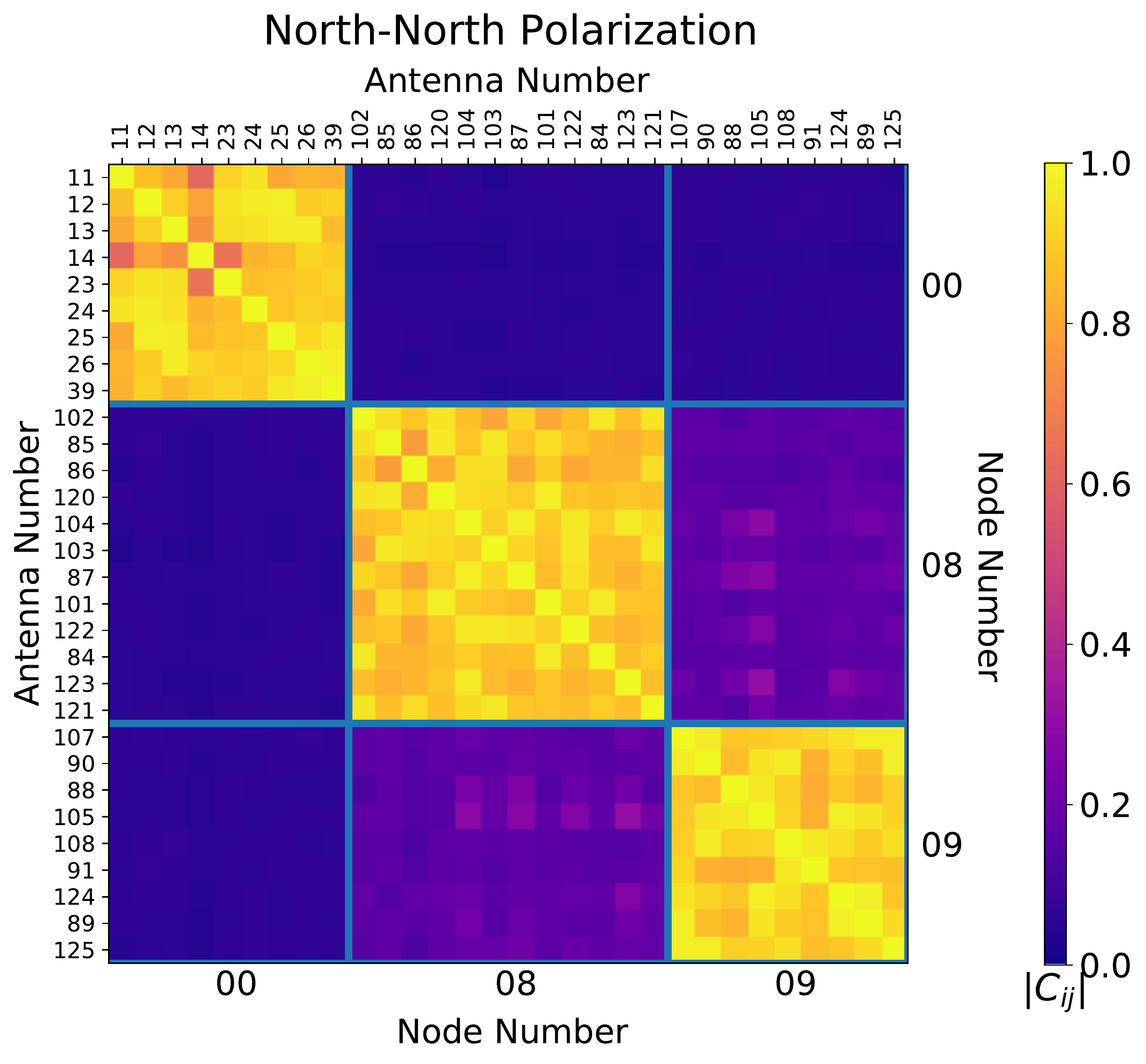}
    \caption{A correlation matrix for a single polarization of HERA phase II data from October 21, 2019 (JD 2458778), taken at a time when the timing system was malfunctioning and antennas between different nodes were not correlating, showing a clear block-diagonal along node lines. This is a sample case where the auto-correlations are nominally acceptable, and investigation of the cross-correlations is necessary to see this type of failure mode.}
    \label{fig:brokenWR}
\end{figure}

\subsubsection{Understanding the Correlation Metric with Simulations}

Figure \ref{fig:corrMatrix} shows that there is a significant amount of structure in the correlation matrices, specifically related to node connections. Baselines within a node appear to have larger values of $C_{ij}$ than baselines between nodes. We have previously noticed instances of severe node-based structure when there are timing mismatches between nodes due to a failure of the clock distribution system. Figure \ref{fig:brokenWR} is an example from an observation when the timing system was known to be broken, and we see clearly that timing mismatches depress the correlation metric. This causes much clearer node structure than the more common structure seen in Figure \ref{fig:corrMatrix}. Therefore, one wonders: are the larger $C_{ij}$ values on the intra-node baselines due to some milder form of this clock distribution issue---perhaps a small error in timing---or is this structure otherwise explicable or even expected? 

Put another way, what is the expectation value of $C_{ij}$ as defined in Equation~\eqref{eq:corr_metric}? We can make the assumption that $\langle V^\text{even}_{ij} \rangle = \langle V^\text{odd}_{ij} \rangle \equiv V^\text{true}_{ij}$ and that the two only differ by their noise, $n_{ij}$, with mean 0 and variance $\sigma_{ij}^2$. Ignoring time and frequency dependence, then we can use Equation~\eqref{eq:corr_metric} to first order (ignoring correlations between the numerator and denominator) to find that
\begin{equation}
    \langle C_{ij} \rangle = \left\langle \frac{(V^\text{true}_{ij} + n^\text{even}_{ij}) (V^\text{true}_{ij} + n^\text{odd}_{ij})^*} {\left|V^\text{true}_{ij} + n^\text{even}_{ij}\right| \left|V^\text{true}_{ij} + n^\text{odd}_{ij}\right|} \right\rangle \approx \frac{\left| V^\text{true}_{ij} \right|^2}{\left| V^\text{true}_{ij} \right|^2 + \sigma_{ij}^2 }.
\end{equation}
This approximate expectation value shows us the importance of the signal-to-noise ratio (SNR). At high SNR, $\langle C_{ij} \rangle$ goes to 1, assuming the two even and odd signal terms are actually the same---i.e.\ that the array is correlating. At low SNR, $\langle C_{ij} \rangle$ goes to 0. 

It follows then that the apparent node-based structure in $C_{ij}$ might actually be the impact of the the relationship between SNR and baseline length. Inspecting the array configuration (see Figure~\ref{fig:array_status}) we see that baselines within the same node tend to be shorter than baselines involving two nodes. Shorter baselines are dominated by diffuse galactic synchrotron emission, which means that they tend to have a higher signal than longer baselines. Since all baselines have similar noise levels and since higher SNR leads to larger values of $C_{ij}$, this could account for the effect. 

\begin{figure*}
    \centering
    \includegraphics[width=1\textwidth]{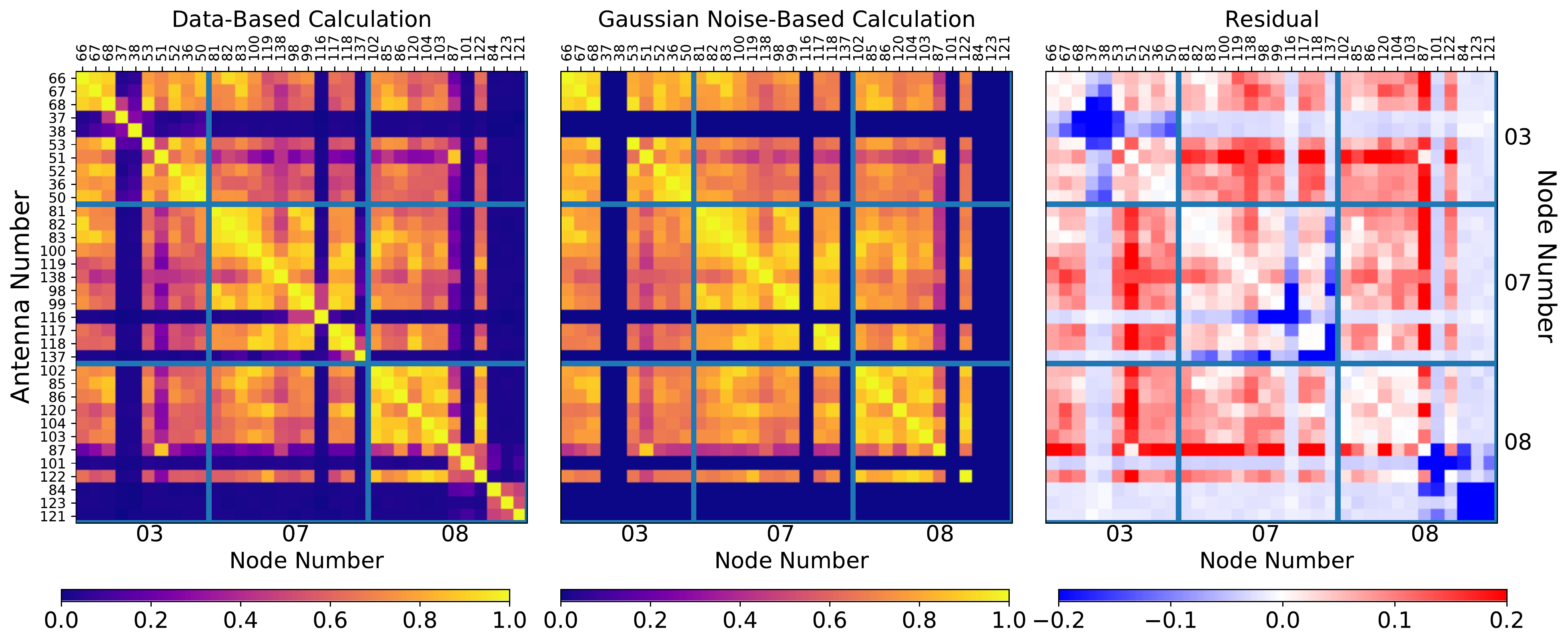}
    \caption{Comparison of the correlation metric computed using true noise from the data (left) and simulated Gaussian thermal noise calculated using the auto-correlations (middle), along with the residual (right). We see here clearly that the node-related structure observed in Figure \ref{fig:crossMatrix} is fully reproduced using simulated Gaussian noise in lieu of the measured noise used in the original calculation.}
    \label{fig:matrixSim}
\end{figure*}

In order to confirm that our node structure is explicable as a baseline length effect rather than some other systematic, we can implement a simple simulation with thermal noise. We calculate $V^\text{true}_{ij}$ from our data as $(V^\text{even}_{ij} +  V^\text{odd}_{ij}) / 2$, and take this as a reasonable stand-in for the sky signal, in lieu of a more sophisticated simulation, since it should have approximately the right relative power and should largely average out the instrumental noise. To each visibility $V^\text{true}_{ij}$ we then add independent Gaussian-distributed thermal noise, with variance given by 


\begin{equation}
    \sigma^2_{ij} = \frac{|V_{ii} V_{jj}|}{\Delta t \Delta \nu},
\end{equation}
where $\Delta t$ is the integration time and $\Delta \nu$ is the channel width. This noise is uncorrelated between baselines, times, and frequencies. We then calculate $C_{ij}$. We compare the $C_{ij}$ with simulated noise to the observed $C_{ij}$ in Figure \ref{fig:matrixSim}. We can see clearly that the node-based structure we observed in the original correlation matrices is completely reproduced when using a Gaussian noise estimate. This conclusion helps confirm that apparent node-based structure in $C_{ij}$ is is driven by sky feature amplitude, which sets the SNR, rather than systematics.

\begin{figure*}
    \centering
    \includegraphics[width=0.7\textwidth]{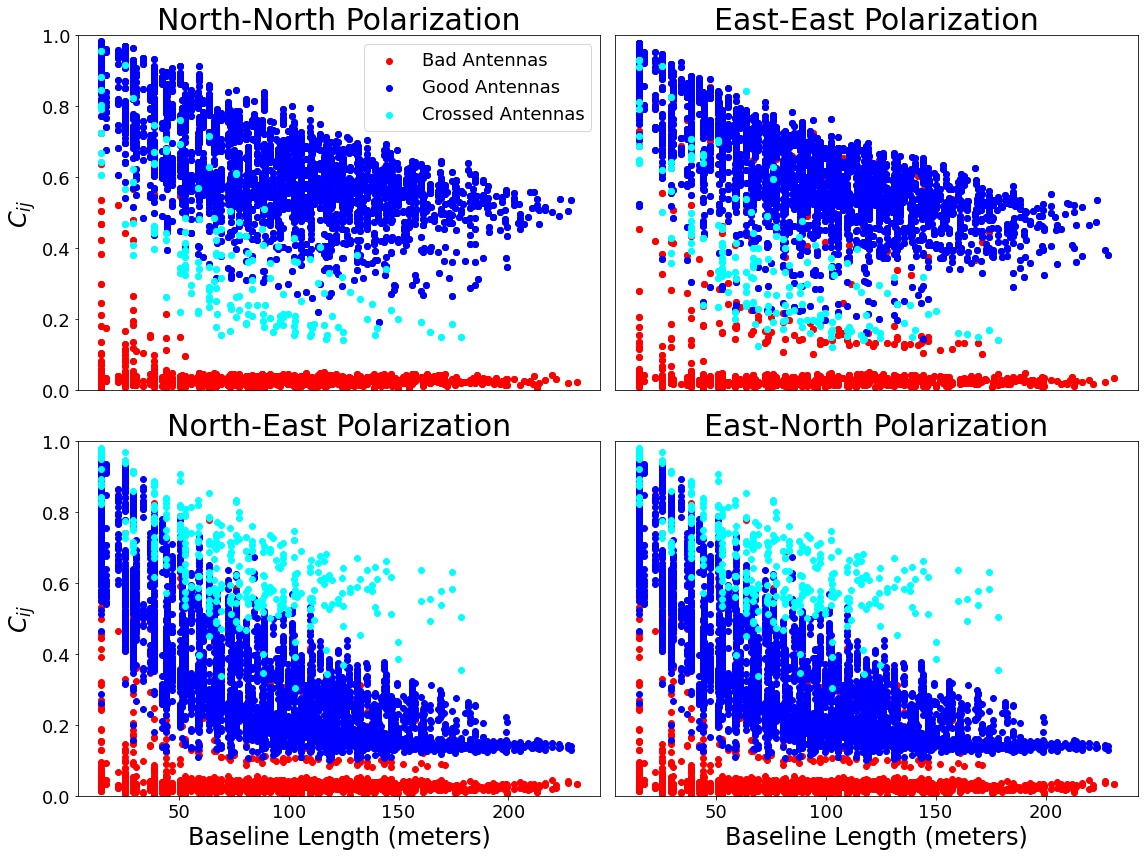}
    \caption{The correlation metric $C_{ij}$ for real data plotted versus baseline length for all four polarizations, with red points representing baselines including at least one antenna that was flagged by this metric, cyan points representing baselines including at least one antenna that was identified as cross-polarized (see section \ref{sec:crossPol}), and blue points representing baselines where neither constituent antenna is flagged. We observe that nominally well-functioning antennas follow an expected power law shape for galactic emission as a function of baseline length, and we note that cross-polarized antennas are clearly identifiable as having excess power in the North-East and East-North polarizations.}
    \label{fig:blLengthSummary}
\end{figure*}

Finally, in Figure \ref{fig:blLengthSummary} we confirm that our metric distribution is representative of the sky by plotting $C_{ij}$ versus baseline length for all four polarizations using real sky data. We color each baseline by whether both constituent antennas were unflagged (blue), at least one was flagged for having a low correlation metric (red), or at least one was flagged for being cross-polarized (cyan). We clearly see the smooth distribution we would expect from sky features, with clearly distinguishable sub-groups by flagging categorization. We would expect a power law slope for galactic emission with strong variation as a function of baseline azimuthal angle, while the point source component should be independent of baseline length or angle, and noise should be similar to point sources \cite{Byrne_2021}. Notably, the nominally good antennas generally follow this pattern, with a strong increase towards shorter baselines. Additionally, baselines observing the North-East and East-North polarizations of the sky show a potential transition between galactic domination to point source or noise domination around 100 meters. At frequencies near the middle of the HERA band this corresponds to 1.5 degrees, which is roughly the scale at which point sources are commensurate with galactic emission \cite{Byrne_2021}. Given the significant agreement between our measured and expected distributions of $C_{ij}$, we are confident in our conclusion that the observed structure in Figure \ref{fig:corrMatrix} is driven by sky features rather than instrumental systematics.

\subsection{Identifying Cross-Polarized Antennas}
\label{sec:crossPol}

As we have already seen in passing, the correlation metric $C_{ij}$ clearly identifies cross-polarized antennas. Here, cross-polarized means that the physical cables carrying the East and North polarization measurements got swapped in the field.  When things are hooked up correctly, we expect to see a stronger correlation between matching polarizations (i.e. $EE$\footnote{HERA dipoles, being fixed, are referred to by their cardinal directions. This avoids much confusion.}  and $NN$), and a weaker correlation between different polarizations. Cross polarized antennas have the opposite situation, with stronger correlation in $EN$ and $NE$. 

We identify this situation automatically with a cross-polarization metric formed from the difference between four polarization combinations in the per-antenna correlation metric: 
\begin{equation}
    C_i^{P_\parallel - P_\times} \equiv \frac{1}{N_\text{ants} - 1} \sum_{j\neq i}{(C_{ij}^{P_\parallel}-C_{ij}^{P_\times})}. \label{eq:CP_i}
\end{equation}
where $P_\parallel$ is either the EE or NN polarization, and $P_\times$ is either the NE or EN polarization.

\begin{figure*}
    \centering
    \includegraphics[width=1.0\textwidth]{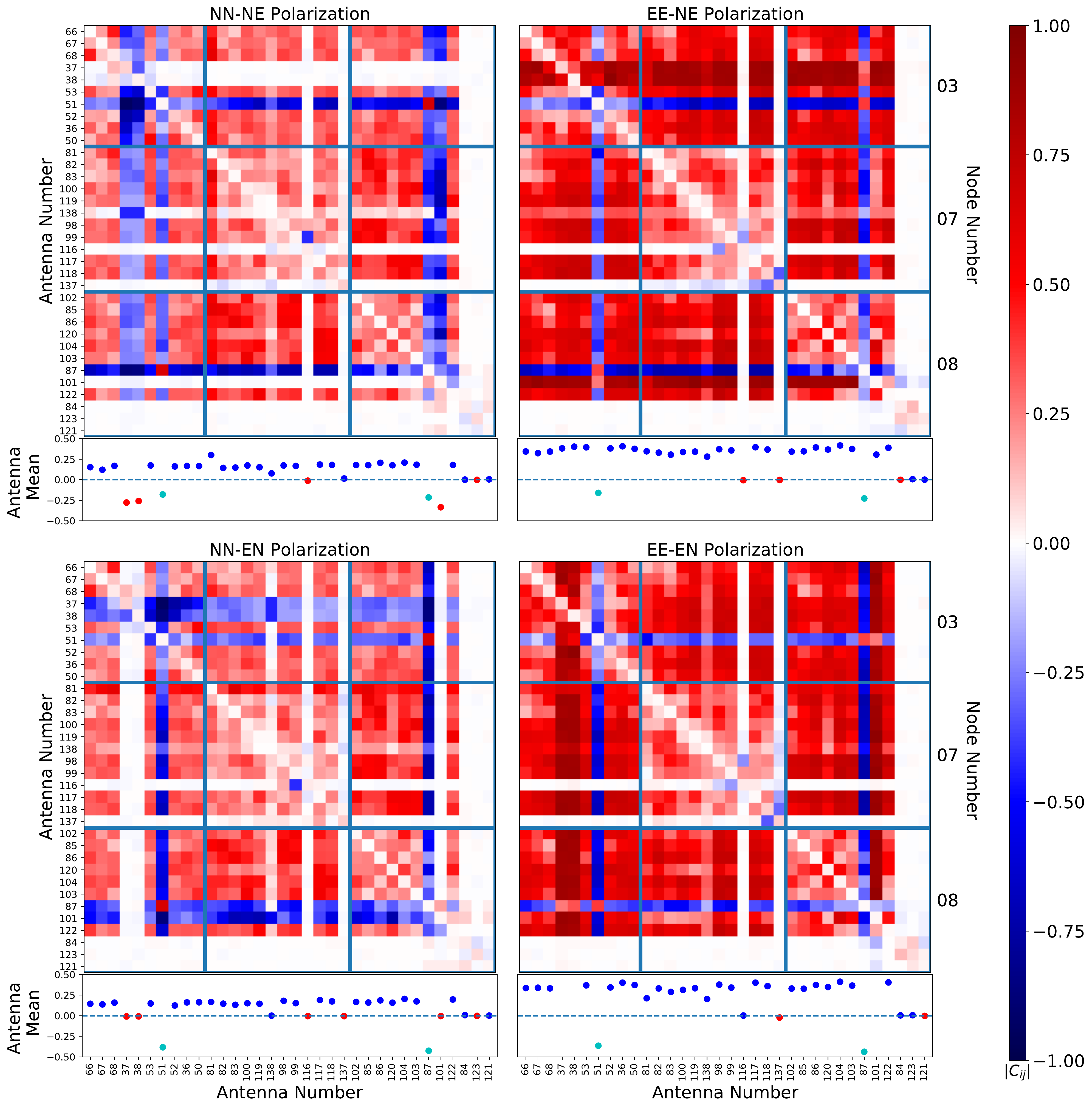}
    \caption{The cross-polarization metric defined in Equation \ref{eq:CP_i}. Any antenna with an average metric $R_i$ that is negative in all four polarization combinations is deemed cross-polarized and marked in the lower panels in cyan. Antennas with a positive antenna mean are marked with blue dots, and those with a negative mean are marked with red dots. Here antennas 51 and 87 are cross-polarized.}
    \label{fig:crossMatrix}
\end{figure*}

We then calculate our cross-polarization metric as the maximum of the four combinations of same-polarization and opposite-polarization visibilities:
\begin{equation}
    R_i = \text{max}\left\{ C_i^{NN-NE}, C_i^{NN-EN}, C_i^{EE-NE}, C_i^{EE-EN} \right\} \label{eq:R_i}
\end{equation}
We take the maximum because it's possible to get negative values for some of the $C_{i}^{P_\parallel - P_\times}$ when one polarization is dead and the other is not. However, when all four values are negative (ie a negative maximum), then the antenna is likely cross-polarized. In Figure~\ref{fig:crossMatrix} we show each of the four differences of $C_{ij}$. Two antennas, 51 and 87, show negative values in all four combinations, indicating swapped cables. Three other antennas---37, 38, and 101---show up negative in two polarizations, which indicate a single dead polarization, rather than a swap.

\subsection{Identifying and Removing Antennas in Practice}
\label{sec:flagging}

Using our correlation metric $C_i$ defined in Equation \ref{eq:C_i} and our cross-polarization statistic $R_i$ defined in Equation \ref{eq:R_i} we can implement an iterative algorithm to flag and remove broken and cross-polarized antennas. In Figure \ref{fig:corrMatrix} we clearly saw that dead antennas have a value of $C_i$ very near zero. As a result, when we calculate $C_i$ for functional antennas by averaging over all constituent baselines, the low correlation between a functional and a dead antenna will decrease the overall value of $C_i$ for the functional antenna. In the case where only a couple of antennas are broken among the whole array this may be tolerable, but it is possible for this bias to cause functional antennas to look much worse than they are, and to potentially drop below the flagging threshold. 

\begin{figure}[H]
    \centering
    \includegraphics[width=1.0\textwidth]{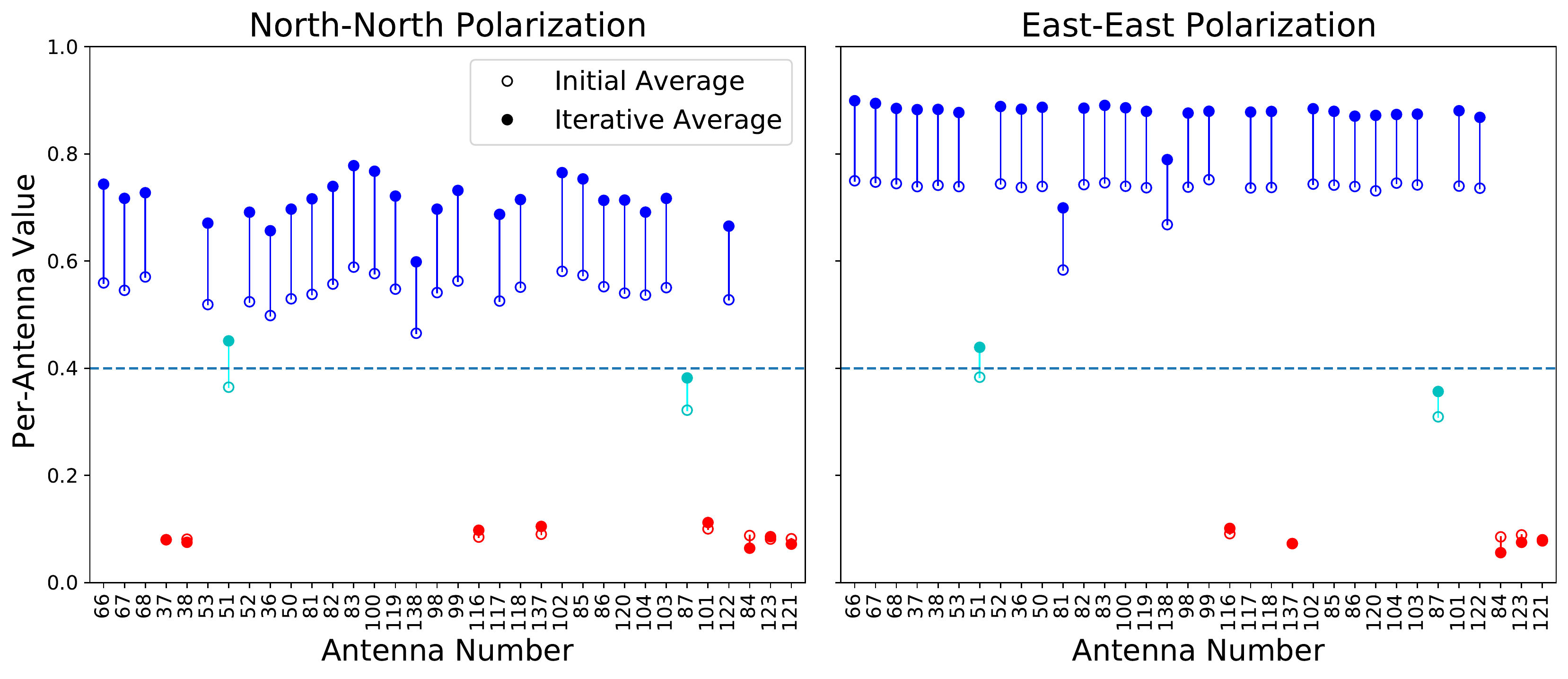}
    \caption{Comparison of the final value of $C_i$ calculated for each antenna using a direct average over $C_{ij}$ versus the iterative calculation outlined in Algorithm \ref{alg:ant_metrics}. We see that using the iterative algorithm helps create a clearer boundary between functional and nonfunctional antennas. Antennas marked in cyan are those that were flagged for being cross-polarized. This plot uses the same representative subset of antennas used in Figures \ref{fig:corrMatrix} and \ref{fig:matrixSim}.}
    \label{fig:directVsIterative}
\end{figure}

To prevent the expected value of our metric from being biased by dead antennas, we implement an iterative metric calculation and flagging approach, outlined in Algorithm \ref{alg:ant_metrics}. First, we calculate $C_i$ for all antennas and identify any that are completely dead (i.e. $C_i$=0) and remove them. Then, recalculate $C_i$ and $R_i$ for all antennas, identify and remove the worst antenna if it falls below the threshold. We continue with this recalculation and reassessment of the metrics until all remaining antennas are above the threshold in both metrics. Figure \ref{fig:directVsIterative} shows a comparison between the values of $C_i$ calculated by directly averaging $C_{ij}$ for each antenna versus using the iterative algorithm. We see clearly from this figure that implementing an iterative approach brings our data into a truly bimodal realm where establishing a threshold is straightforward. Based on the observed values, we set an empirical threshold of $C_i=0.4$, such that any antennas below that value will be flagged and removed. Note that the two antennas marked in cyan are both cross-polarized, so their value near the threshold is not worrisome. As noted in section \ref{sec:crossPol}, these points are flagged for having a maximum value of $R_i$ below zero. This iterative approach to flagging is robust against varying proportions of broken antennas, which is essential for flagging during the commissioning phase of an array. While the iterative approach somewhat increases computation time, we find the trade-off to be worthwhile. Even with the iterative approach, flagging based on the cross-correlation metrics scales at worst with the number of visibilities, which we find reasonable. Our most computationally expensive step is simply reading in all of the data. In the case where this step becomes computationally prohibitive in a real-time pipeline, we may take advantage of the time-stability of the correlation metric and calculate antenna flags on a sparser time interval. As it is computationally infeasible to hold data for all baselines over the whole night in memory at once, we reserve time domain data quality assessments for auto-correlations only, as discussed in Section \ref{ch:auto_metrics}.

It is also relevant to note that we only use the North-North and East-East polarizations during antenna flagging. There are multiple reasons for this. First, we see in Figure \ref{fig:blLengthSummary} that the expected value of $C_{ij}$ is higher in same-polarization correlations compared to cross-polarization correlations. The larger separation in expected value between functional and dead antennas leads to a more robust flagging threshold. Second, we have no evidence in HERA data to indicate the existence of systematics that appear only in correlations between different polarizations. Therefore, flagging only on the same-polarizations allows for clearer distinction between functional and broken antennas without missing any known failure modes of the system.

\section{Auto-Correlation Metrics}
\label{ch:auto_metrics}

While the correlation metrics provide an absolute check on data quality of a particular antenna, not all effects will be caught by this approach. For example, if one antenna has a bandpass structure completely unlike the rest---an effect that might be calibratable---it is useful to identify it and flag it as a symptom of some deeper malfunction in the array. It is useful, therefore, to assess antennas for ways in which they deviate from others, assuming that the plurality of antennas will be well-behaved.\footnote{Even when the majority of antennas are malfunctioning, our iterative techniques for outlier detection can still be robust when the malfunctions are multi-causal. To crib from \emph{Anna Karenina}, all happy antennas are alike, but every unhappy antenna is unhappy in its own way.} 

Identification of misbehavior is more difficult with a new system. A newly-built telescope system with novel combinations of technologies means that we lack an a-priori model for how signal chains might malfunction. In early commissioning we observed broadband temporal and spectral instabilities in visibilities which motivated a metric that examines whole nights of data. 

We choose to focus on auto-correlations $V_{ii}$ for two reasons. The first is data volume. The number of auto-correlations scales with $N_\text{ant}$ while the number of visibilities scales with $N_\text{ant}^2$---far too big to load into memory at once for a whole night of data. Second, because our goal is to identify malfunctioning antennas before calibration, we focus on auto-correlations because they are easier to compare without calibration. Comparison between visibilities measuring the same baseline separation requires at minimum a per-antenna delay calibration to flatten phases. That term in autocorrelations cancels out, leaving each $V_{ii}^\text{obs} \propto |g_i|^2 V_\text{auto}^\text{true}$. Since most bandpass gains should be similar, auto-correlations can be sensibly compared to one other to look for outliers before calibrating. Even if $|g_i|^2$ differs between antennas, that is something we would like to know and perhaps rectify in the field.

Historically, auto-correlations from radio interferometers are seldom used. For example, at the VLA the autos are usually discarded \cite{Taylor1999}. The usual reasons given for this are that auto-correlations have a noise bias and that gain variations are assumed to not correlate between antennas. However, 
given HERA's sensitivity to calibration stability, this assumption is worth re-considering. Recently, other collaborations have also begun exploring auto-correlations as a valuable tool for assessing data quality \cite{Rahimi} and performing calibration \cite{Barry_2019}.


Each antenna's auto-correlation stream can be reduced statistically across an entire observation to a single metric spectrum which can then be quickly compared to all other spectra to search for outliers. For HERA, a drift-scan telescope which operates continuously each night for months at a time, one full night's observation time is a useful averaging time range. We focus on four factors motivated by antenna failure modes noted in manual inspection of hundreds of antenna-nights of autocorrelation data: bandpass shape (Section~\ref{sec:shape}), overall power level (Section~\ref{sec:power}), temporal variability (Section~\ref{sec:temp_var}), and temporal discontinuities (Section~\ref{sec:temp_discon}). The purpose of this section is to develop quantitative metrics that capture these qualitative concerns in a rigorous way, attempting to reduce antenna ``badness'' along each of these dimensions to a single number.  In Section~\ref{sec:practice} we show how these four statistics together produce a useful summary of per-antenna performance (see Figure~\ref{fig:dashboard}).

Each of these four statistics comes in two flavors. The first is a median-based statistic which is more robust against transient or narrow-band outliers in each time vs.\ frequency plot or ``waterfall", like RFI. The second is a more sensitive mean-based statistic. Our basic approach, outlined in pseudocode in Algorithm~\ref{alg:auto_metrics}, is to remove the worst antennas with the robust statistics, then flag RFI, then flag the more subtly bad antennas with the mean-based statistics. In the following sections, we offer a more precise definition of the calculations and the algorithmic application.

\subsection{Outliers in Bandpass Shape} \label{sec:shape}

Our first metric is designed to identify and flag antennas with discrepant bandpass structures. This often indicates a problem in the analog signal chain. As we mention in Algorithm~\ref{alg:auto_metrics}, we first reduce the auto-correlation for antenna $i$, polarization $p$ to a single spectrum $S(\nu)$ as follows.
\begin{equation}
    S^\text{med}_{i,p}(\nu) \equiv \frac{\med{V_{ii,pp}(t,\nu)}_t}{\med{V_{ii,pp}(t,\nu)}_{t,\nu}} 
    \label{eq:med_shape_spec}
\end{equation}
where $\med{}_t$ indicates a median over time while $\med{}_{t,\nu}$ indicates a median over both time and frequency. This gives us a notion of the average bandpass shape while normalizing the result to remove differences between antennas due to overall power. The reduction from waterfall to spectrum only needs to be computed once per antenna.

We can now compute the median difference between each antenna's spectrum and the median spectrum with the same polarization $p$ according to the following formula:
\begin{equation}
    D^\text{med}_{i,p} \equiv \med{\left|S^\text{med}_{i,p}(\nu) - \med{ S^\text{med}_{j,p}(\nu) }_j \right|}_{\nu},
    \label{eq:med_shape_dist}
\end{equation}
where $j$ indexes over all unflagged antennas. To determine which antenna to flag, if any, we convert each $D^\text{med}_{i,p}$ into a modified $z$-score by comparing it to the overall distribution of distances. These modified $z$-scores are defined as
\begin{align}
    z^\text{mod}_{i,p} & \equiv \frac{\sqrt{2}\text{erf}^{-1}(0.5)\left( D_{i,p} - \med{ D_{j,p} }_j \right)}{\MAD{D_{j,p} }_j} 
    \nonumber \\ 
    & \approx 0.67449 \left(\frac{ D_{i,p} - \med{ D_{j,p} }_j}{\MAD{D_{j,p} }_j}\right), \label{eq:modz} 
\end{align}
where $\MAD{}_j$ is the median absolute deviation over antennas and $\text{erf}^{-1}(x)$ is the inverse error function. The factor of $\sqrt{2}\text{erf}^{-1}(0.5)$ normalizes the modified $z$-score so that the expectation value of a $z^\text{mod}$ of a sample drawn from a Gaussian distribution is the same as its standard $z$-score.\footnote{Were the distribution of distance metrics Gaussian (it is generally not), then one could think of modified $z$-score of 8 as an ``8$\sigma$ outlier.'' This kind of language is imprecise, but often useful for building intuition.}

Having computed modified $z$-scores for every antenna and every polarization, we iteratively remove the antenna with the worst modified $z$ over all metrics and both polarizations. When one polarization is flagged, we flag the whole antenna. We then recompute $D^\text{med}_{i,p}$ and $z^\text{mod}_{i,p}$ and continue flagging antennas until none have a modified $z$-score over a chosen threshold, in our case 8.0. All subsequent metrics use the same threshold for median-based flagging.

\begin{figure*}[h]
    \centering
    \includegraphics[width=1.0\textwidth]{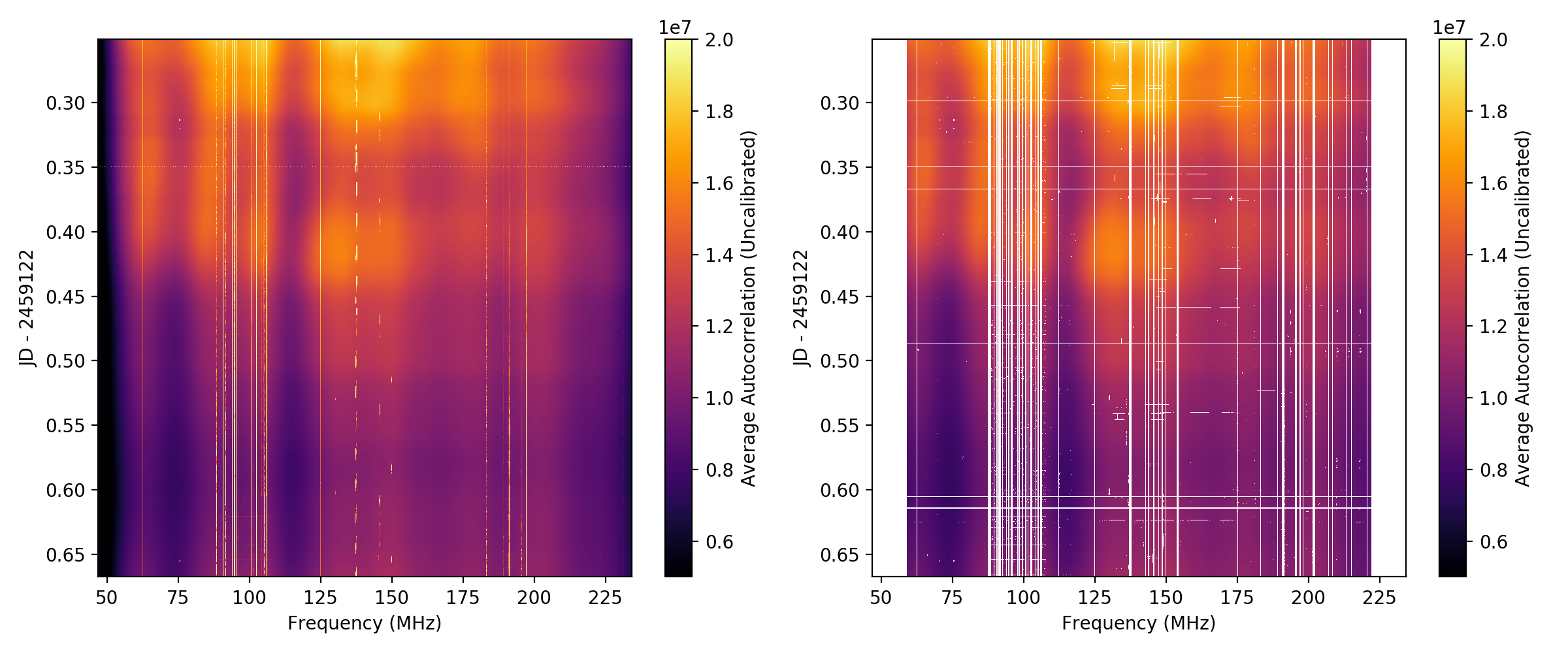}
    \caption{Auto-correlation averaged over good antennas, before and after RFI flags. RFI is excised using local median and mean filters to search for outliers, followed by 1D thresholding. This is a simplified version of the algorithm used in \citeA{H1C_IDR2_Limit} with the exception that it is sped up by performing it on a single waterfall averaged over unflagged antennas.}
    \label{fig:avg_auto_flagged}
\end{figure*}

Next we perform a simple RFI flagging, analogous to the algorithm used in \citeA{H1C_IDR2_Limit}, but performed on a single  auto-correlation waterfall averaged over all remaining antennas. This process includes a search for local outliers after median filtering and then mean filtering, which are flagged as RFI. Finally, a thresholding algorithm is performed that throws out entire channels or entire integrations which are themselves significant outliers after analogous 1D filtering. The results of this process are shown in Figure~\ref{fig:avg_auto_flagged}. This process flags 12.6\% of the data, excluding band-edges, and leaves 11.3\% of channels and 1.0\% of all times completely flagged. This is likely an under-count of RFI; the algorithm is to designed to flag the most egregious outliers that might skew the statistics described below, rather than to find and remove RFI for the purpose of making sensitive 21\,cm power spectrum measurements.

After RFI flagging, we next compute shape metric spectra with mean-based statistics. Analogously to Equation~\ref{eq:med_shape_spec} this case, 
\begin{equation}
    S^\text{mean}_{i,p}(\nu) \equiv \frac{\mean{V_{ii,pp}(t,\nu)}_t}{\mean{V_{ii,pp}(t,\nu)}_{t,\nu}}, \label{eq:mean_shape_spec}
\end{equation}
where $\mean{}_t$ indicates a weighted-mean over the time dimension, giving zero weight to times and frequencies flagged for RFI. Likewise, these spectra are reduced to scalar distance metrics as
\begin{equation}
    D^\text{mean}_{i,p} \equiv \mean{\left|S^\text{mean}_{i,p}(\nu) - \mean{ S^\text{mean}_{j,p}(\nu) }_j \right|}_{\nu},
    \label{eq:mean_shape_dist}    
\end{equation}
where again averages are performed over unflagged antennas, times, and frequencies. Just as before, we compute modified $z$-scores to iteratively flag the worst antenna outlier, recalculating $D^\text{mean}_{i,p}$ after each antenna is flagged. This proceeds until no antennas exceed a $z$-score of 4; half that used during the first round median cut. Again, this threshold is the same for mean-based flagging in all subsequent metrics.

\begin{figure*}[h]
    \centering
    \includegraphics[width=1.0\textwidth]{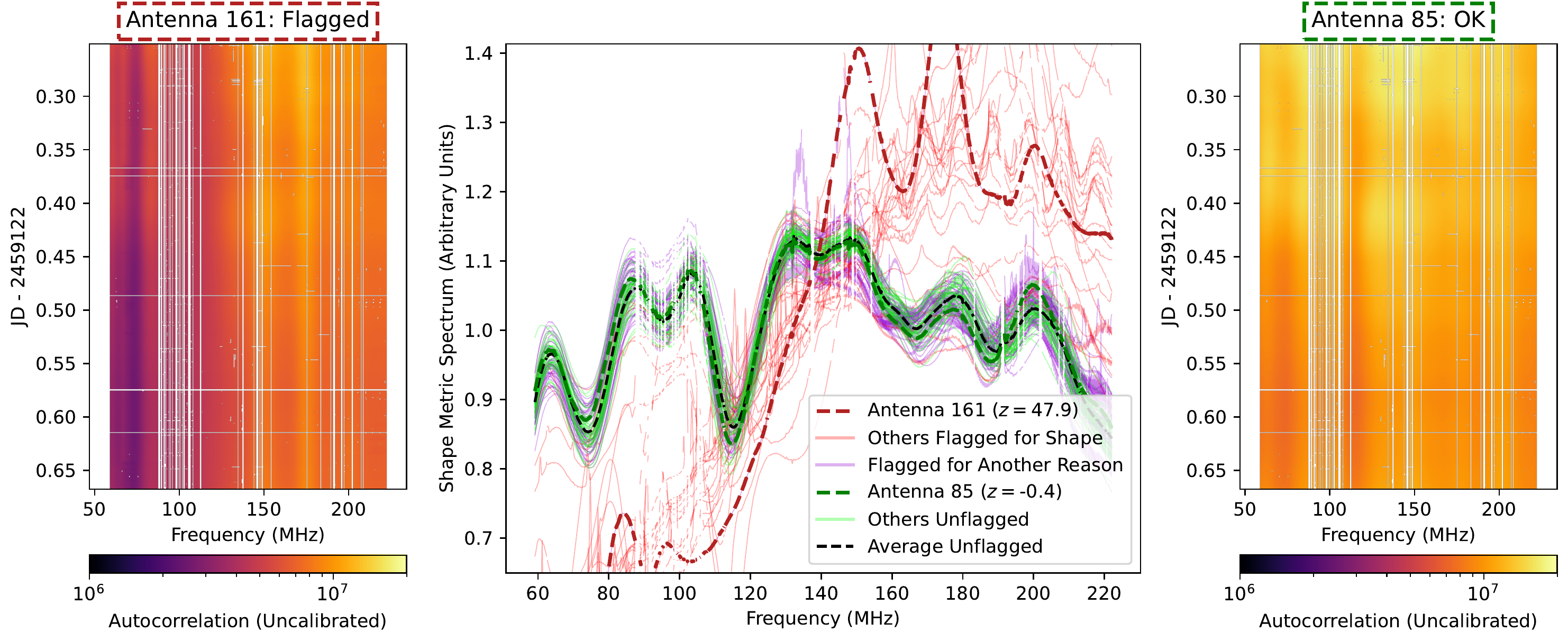}
    \caption{Here we show the shape metric spectra, defined in Equation~\ref{eq:mean_shape_spec}, for all North/South-polarized antennas in the array (center panel). Outliers (red lines) are defined as has having a modified $z$-score greater than 4.0 in their scalar distance metric (Equation~\ref{eq:mean_shape_dist}) compared compared to the average good antenna (black dashed line) and the distribution of good antennas (light green lines). Note that this figure includes flagging by three other metrics causing some antennas to be flagged even though they look okay here.  We highlight two example antennas and show their full auto-correlation waterfalls, one flagged (161; left panel and dark red dashed line) and one functioning normally (85; right panel and dark green dashed line).}
    \label{fig:shape_comparison}
\end{figure*}
In Figure~\ref{fig:shape_comparison} we show the the results of this operation with example waterfalls and metric spectra for antennas that were and were not flagged by our modified $z$-score cut of 4.0. In general, we find that the metric robustly identifies antennas with metric spectra discrepant from the main group of antennas. Almost everything in red in Figure~\ref{fig:shape_comparison} is a pretty clear outlier. Where exactly to draw the line is tricky, and likely requires some manual inspection of metric spectra and waterfalls for antennas near the cutoff. Note that this figure includes flagging by all four metrics. Some moderate outliers in shape were not flagged for shape but were flagged for other reasons, indicating that this metric and the other three discussed below are not completely independent.

\subsection{Outliers in Bandpass Power}\label{sec:power}

We next turn to looking for outliers in bandpass power. High power might indicate incorrect amplifier settings while a signal chain malfunction might cause anomalously low power.  Our approach for finding outliers in power is very similar to the one for finding outliers in bandpass shape laid out in Section~\ref{sec:shape}. Here we lay out the mathematical approach, highlighting and motivating differences between the two.

Once again, we begin by defining median-based metric spectra which collapse each antenna's waterfall down to a single number per frequency. For bandpass power, that is simply
\begin{equation}
    S^\text{med}_{i,p}(\nu) \equiv \med{V_{ii,pp}(t,\nu)}_t.
    \label{eq:med_power_spec}
\end{equation}
This is simply an unnormalized version of Equation~\ref{eq:med_shape_spec}. However, instead of directly comparing each antenna's spectrum with the median spectrum, we instead compare their logarithms:
\begin{align}
    D^\text{med}_{i,p} \equiv \text{med} \Big\{ \Big| 
    & \log\left(S^\text{med}_{i,p}(\nu)\right) -  \nonumber \\ 
    &  \log\left(\med{ S^\text{med}_{j,p}(\nu) }_j \right) \Big| \Big\}_{\nu},
    \label{eq:med_power_dist}
\end{align}
This logarithmic distance measure reflects the fact that gains are multiplicative and that the optimal ranges for amplifier and digitization are themselves defined in decibels.
We take the absolute value of the difference of the logs because we want to penalize both antennas with too little power, which may indicate a malfunction, and antennas with too much power, which may cause a nonlinear response to the sky signal. 

After RFI flagging as described in the previous section, we next proceed with outlier detection using modified mean-based statistics, which are straightforward adaptations of Equations~\ref{eq:med_power_spec} and \ref{eq:med_power_dist}:
\begin{equation}
    S^\text{mean}_{i,p}(\nu) \equiv \mean{V_{ii,pp}(t,\nu)}_t,
    \label{eq:mean_power_spec}
\end{equation}
\begin{equation}
    D^\text{mean}_{i,p} \equiv \mean{\left|\log\left(S^\text{mean}_{i,p}(\nu)\right) - \log\left(\mean{ S^\text{mean}_{j,p}(\nu) }_j \right) \right|}_{\nu}.
    \label{eq:mean_power_dist}
\end{equation}

\begin{figure*}[h]
    \centering
    \includegraphics[width=1.0\textwidth]{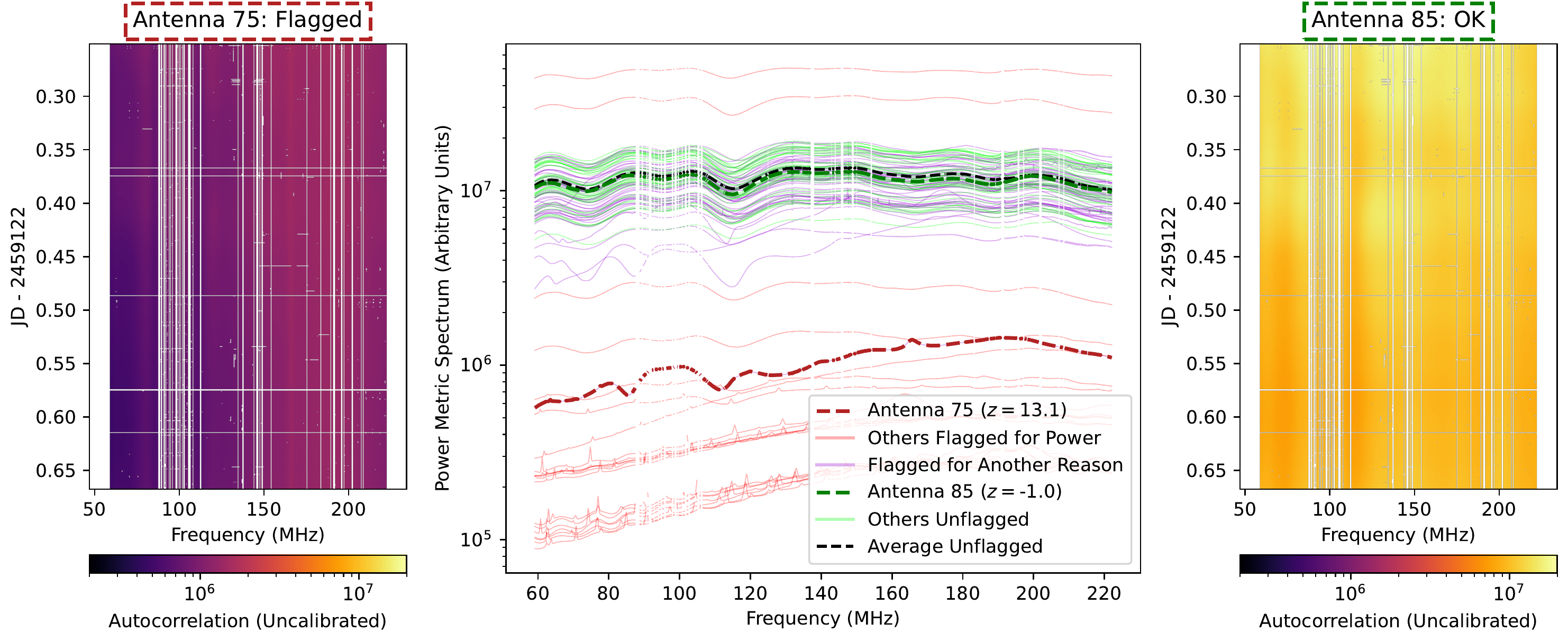}
    \caption{Here we show bandpass power metric spectra, defined in Equation~\ref{eq:mean_power_spec}, for all North/South-polarized antennas in the array (center panel). Just as in Figure~\ref{fig:shape_comparison} we show flagged and unflagged antennas, highlighting example auto-correlation waterfalls of good (85; right panel) and bad (75; left panel) antennas, as defined by the modified $z$-score of their distance metric (Equation~\ref{eq:mean_power_dist}). While antenna 75's bandpass structure is similar to the normal antennas, its autocorrelation has roughly an order of magnitude less power. This makes us suspicious that the amplifiers in the signal chain are not operating properly.}
    \label{fig:power_comparison}
\end{figure*}

Once again, as we can see in Figure~\ref{fig:power_comparison}, this metric picks a number of antennas that are clearly behaving differently than the main group. 
As we saw in the previous section we see there are some antenna which appear to be ``in family'' according to this metric but are flagged for other reasons.  But now we can start to see why this might be. A few of the flagged antennas appear to be fine according to their bandpass shape but are significantly lower or higher in power than the rest.

\subsection{Outliers in Temporal Variability} \label{sec:temp_var}

We now turn to the the question of searching for outliers in the \emph{temporal structure} of the antenna response. While the metrics follow a similar pattern---median-based spectra and distances, followed by mean-based spectra and distances---they are mathematically quite different from those in Sections~\ref{sec:shape} and \ref{sec:power}. 

During observing and subsequent inspection analysis sharp discontinuities were observed in the auto-correlations. Often, though not always, these are rapid changes occurring within a single integration. Sometimes they are accompanied with apparent changes in the bandpass shape or power. Sometimes the effects are relatively localized in frequency; sometimes they are broadband. Sometimes they are frequent jumps; sometimes there are just a handful of discontinuities followed by minutes or hours of stability. Developing a physical understanding of the origin of these effects is an ongoing research effort outside the scope of this paper. Absent that understanding---and a hardware fix to prevent the effects---we have to consider this behavior suspicious and therefore meriting flagging.  

Here and in Section~\ref{sec:temp_discon} we present two metrics for automatically identifying temporal effects. In general, we are looking for forms of temporal structure of the auto-correlations that cannot be explained by the sky transiting overhead. The first looks for high levels of temporal variability throughout the night. To distinguish temporal variability due to sky-rotation from anomalous temporal structure, our metrics are based on a comparison of each antenna's auto-correlation waterfall with an average waterfall over all antennas. For our first round of median statistics, we use the median absolute deviation of the waterfall along the time axis after dividing out the median waterfall over antennas. Thus, 
\begin{equation}
    S^\text{med}_{i,p}(\nu) \equiv \MAD{\frac{V_{ii,pp}(t,\nu)}{\med{V_{jj,pp}(t,\nu)}_j}}_t.
    \label{eq:med_var_spec}
\end{equation}
to produce a single spectrum for each antenna that can be reasonably interpreted as the standard deviation over time of each channel with respect to the mean over time.

We calculate the distance metric for each antenna by taking the median over frequency of how much the antenna's temporal variability metric spectrum exceeds the median metric spectrum over all antennas:
\begin{equation}
    D^\text{med}_{i,p} \equiv \med{S^\text{med}_{i,p}(\nu) - \med{ S^\text{med}_{j,p}(\nu) }_j}_{\nu}.
    \label{eq:med_var_dist}
\end{equation}
Note that we do not take the absolute value of the difference; while shape and power mismatches are penalized both for being too low and for being too high, we do not penalize antennas for varying less that the median. These simply become negative $z$-scores --indicating that an antenna has less temporal variation than the median signal-- and do not result in flags. 

Our mean-based metrics are a straightforward adaptation of Equations~\ref{eq:med_var_spec} and \ref{eq:med_var_dist}:  
\begin{align}
    S^\text{mean}_{i,p}(\nu) \equiv \Bigg[ 
     \mean{\left(\frac{V_{ii,pp}(t,\nu)}{\mean{V_{jj,pp}(t,\nu)}_j}\right)^2}_t & - \nonumber \\   \mean{\frac{V_{ii,pp}(t,\nu)}{\mean{V_{jj,pp}(t,\nu)}_j}}^2_t & \Bigg]^{1/2},
    \label{eq:mean_var_spec}
\end{align}
\begin{equation}
    D^\text{mean}_{i,p} \equiv \mean{S^\text{mean}_{i,p}(\nu) - \mean{ S^\text{mean}_{j,p}(\nu) }_j}_{\nu}.
    \label{eq:mean_var_dist}
\end{equation}

In theory, the denominator of Equations~\ref{eq:med_var_spec} and \ref{eq:mean_var_spec} should change each time an antenna is thrown out and the distance measures and modified $z$-scores are recomputed. This can be computationally expensive when a large fraction of the array needs flagging, as has sometimes been the case during HERA commissioning. In practice, we take a shortcut. During the median-statistics round, we simply neglect this effect, relying on the fact that the median statistics are relatively insensitive to the set of antennas that are flagged. During the next round using mean-based statistics, we iteratively remove antennas until no antennas remain above our modified $z$-score cut. Only then do we recompute the metric spectra in Equation~\ref{eq:mean_var_spec}. In general, this has the effect of making the metric spectra more sensitive to temporal variability, since the mean spectrum will include fewer anomalously variable antennas. The standard procedure of removing antennas and recalculating each $D_{i,p}^\text{mean}$ (but not each $S^\text{mean}_{i,p}(\nu)$) is repeated. This loop continues until no more more antennas are flagged after recalculating $S^\text{mean}_{i,p}(\nu)$ one final time.

As a brief aside, we present Table \ref{table:FlagNums}, which shows the number of antennas flagged by each metric at each step. The table shows that the power and shape metrics are relatively bimodal, in that the vast majority of antennas flagged by those metrics were bad enough to be flagged by the median-based statistics, and very few antennas required the more sensitive iterative approach. In contrast, we see that antennas flagged by the temporal variability and temporal discontinuities (outlined in the next section) metrics have a more gradual distribution of badness, rendering the mean-based iterative flagging step all the more necessary.

\begin{table}[h!]
\centering
\begin{tabular}{ c c c c c c }
\hline
  & Power & Shape & Temporal  & Temporal & Total \\
  & & & Variability & Discontinuties &  \\
\hline
\multirow{3}{4em}{Round 1 (Median-Based)} & & & & & \\
& 19 & 24 & 26 & 19 & 36\\
& & & & & \\
\multirow{3}{4em}{Round 2 (Mean-Based)} & & & & & \\
& 3 & 3 & 13 & 33 & 30 \\
& & & & & \\
\hline
\end{tabular}
\caption{Table showing the number of antennas flagged at each step and for each metric of \texttt{auto\_metrics}. Note that each antenna can be flagged by multiple metrics, so the total number of antennas flagged per round is less than the sum of flags per metric. Additionally, antennas may be flagged for one reason during the median-based round and another during the mean-based round, which explains why the total for each metric can possibly exceed the total for the whole round.}
\label{table:FlagNums}
\end{table}

\begin{figure*}[h]
    \centering
    \includegraphics[width=1.0\textwidth]{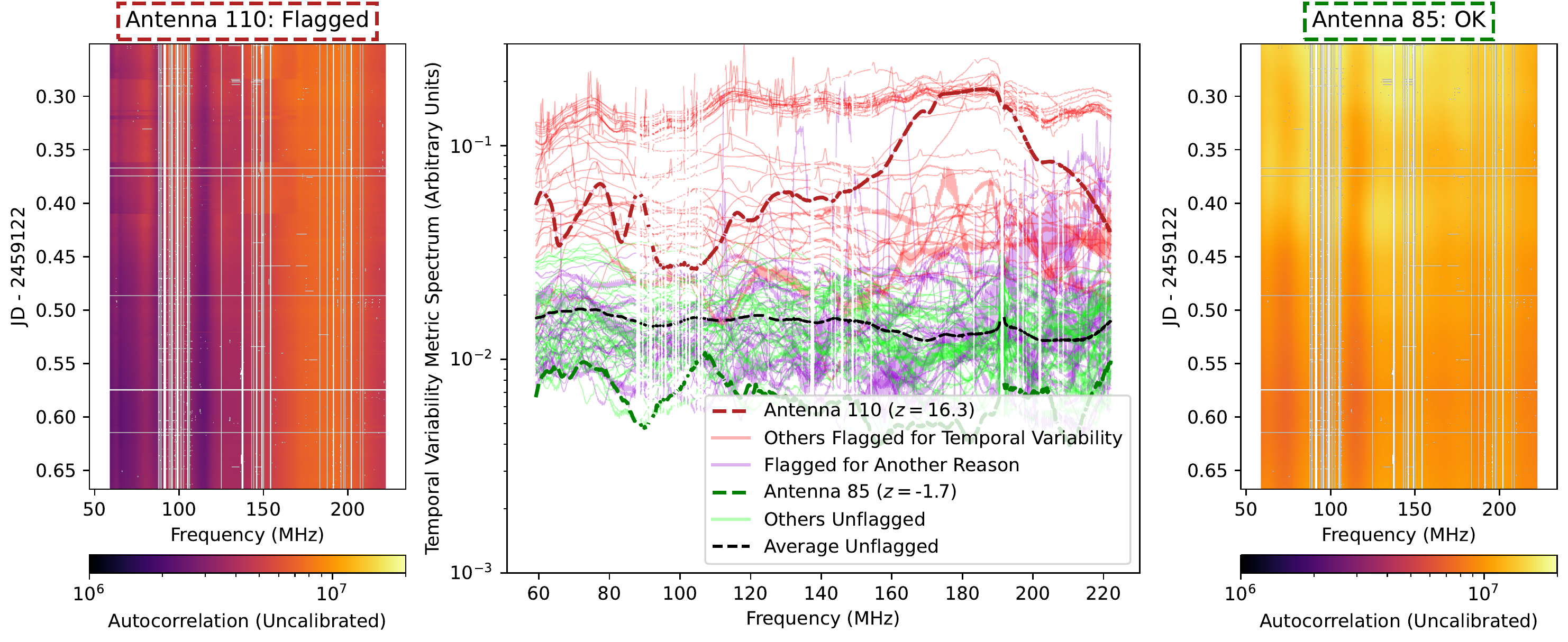}
    \caption{Here we show temporal variability metric spectra, defined in Equation~\ref{eq:mean_var_spec}, for all North/South-polarized antennas in the array (center panel). Just as in Figure~\ref{fig:shape_comparison} we show flagged and unflagged antennas, highlighting example auto-correlation waterfalls of good (85; right panel) and bad (110; left panel) antennas, as defined by the modified $z$-score of their distance metric (Equation~\ref{eq:mean_var_dist}). The malfunction in antenna 110---chunks of time where the waterfall shape and amplitude varies discontinuously---is subtle. It is easiest to see in the waterfall at low frequencies during the first half of the night. These sorts of effects are often more visible in metric spectra and in renormalized waterfalls, as demonstrated in Figure~\ref{fig:dashboard}.}
    \label{fig:variability_comparison}
\end{figure*}

In Figure~\ref{fig:variability_comparison} we show the the resulting mean-based metric spectra after iteratively removing outliers. 
While there are some very clear outliers that are successfully identified, the precise line between what should be considered good and what should be considered bad is ambiguous. Clearly the pathology seen in Antenna 110 is worthy of flagging and the metric successfully identifies it as having high variability relative to the average waterfall. Likewise, most of what is identified as good appears to be behaving like most of the other antennas. Just as with the previous metrics, some level of inspection of antennas near the cutoff is warranted. 

\subsection{Outliers in Temporal Discontinuities} \label{sec:temp_discon}

Though a range of temporal variation pathologies were noted during the observing and data inspection phase one that stood out was abrupt changes occurring faster than the integration time and lasting minutes to hours.  Our second metric for anomalous temporal structure looks for such sharp discontinuities, which also cannot be explained by sky rotation. As with our metric for overall temporal variability (see Section~\ref{sec:temp_var}), our metric is based on examining each antenna's waterfall after dividing out the average waterfall of unflagged antennas. Instead of using the median absolute deviation or the standard deviation, which are measures of variability on any timescale, we instead want to detect variability on the shortest timescale---which is the hardest to explain with antenna-to-antenna primary beam variations \cite{Dillon_2020}. 

Beginning with the auto-correlation scaled by the median over antennas, we compute the discrete difference along the time axis, and then collapse that waterfall (which is only one integration shorter than the original) along the time axis to a metric spectrum. In our first round of flagging using median statistics, this becomes:
\begin{align}
    S^\text{med}_{i,p}(\nu) \equiv \text{med} \Bigg\{ \Bigg| 
    & \frac{V_{ii,pp}(t+\Delta t,\nu)}{\med{V_{jj,pp}(t+\Delta t,\nu)}_j} - \nonumber \\
    & \frac{V_{ii,pp}(t,\nu)}{\med{V_{jj,pp}(t,\nu)}_j} \Bigg| \Bigg\}_t,
    \label{eq:med_discon_spec}
\end{align}
where $\Delta t$ is our integration time (9.6\,s in this data set). Our distance measure, designed to penalize only excessive levels of temporal discontinuities, is the same as in Equation \ref{eq:med_var_dist}:
\begin{equation}
    D^\text{med}_{i,p} \equiv \med{S^\text{med}_{i,p}(\nu) - \med{ S^\text{med}_{j,p}(\nu) }_j}_{\nu}.
    \label{eq:med_discon_dist}
\end{equation}
The adaption to mean-based statistics is straightforward:
\begin{align}
    S^\text{mean}_{i,p}(\nu) \equiv \Bigg< \Bigg| 
    & \frac{V_{ii,pp}(t+\Delta t,\nu)}{\mean{V_{jj,pp}(t+\Delta t,\nu)}_j} - \nonumber \\
    & \frac{V_{ii,pp}(t,\nu)}{\mean{V_{jj,pp}(t,\nu)}_j}\Bigg| \Bigg>_t,
    \label{eq:mean_discon_spec}
\end{align}
\begin{equation}
    D^\text{mean}_{i,p} \equiv \mean{S^\text{mean}_{i,p}(\nu) - \mean{ S^\text{mean}_{j,p}(\nu) }_j}_{\nu}.
    \label{eq:mean_discon_dist}
\end{equation}

\begin{figure*}[h]
    \centering
    \includegraphics[width=1.0\textwidth]{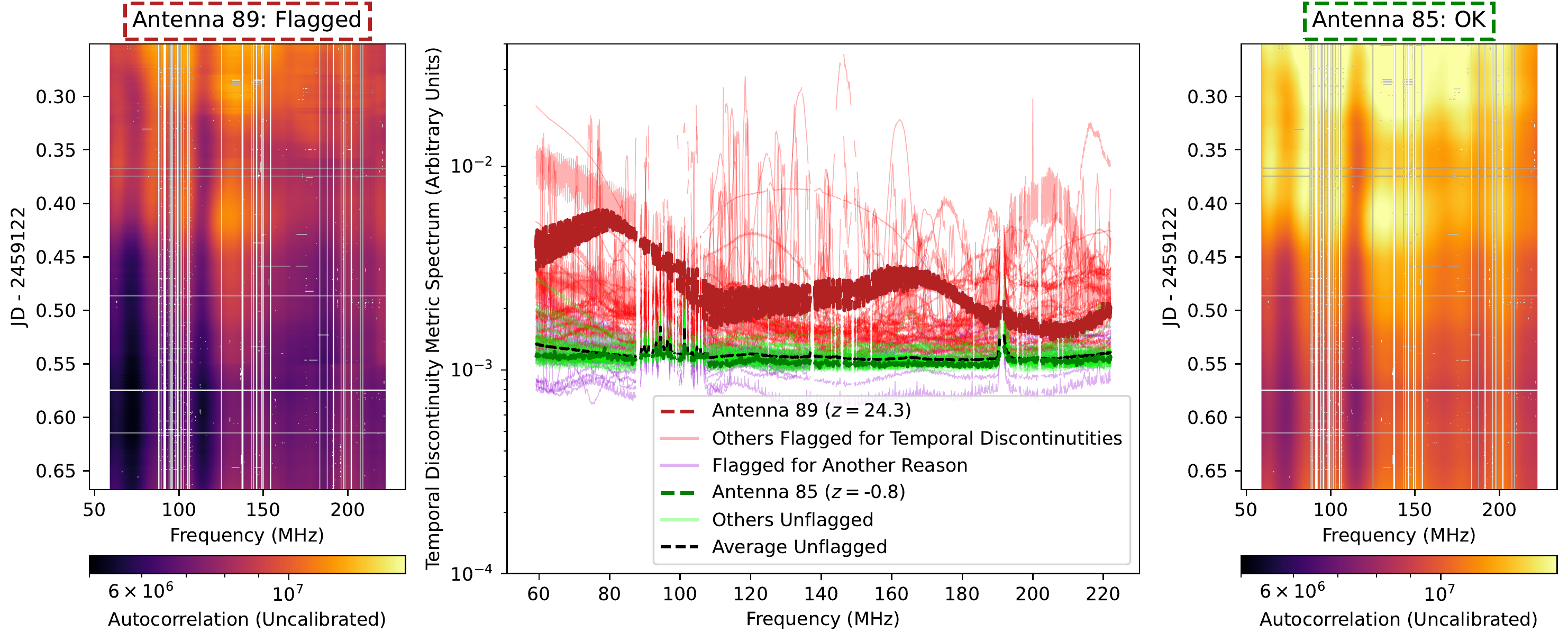}
    \caption{Here we show temporal discontinuity metric spectra, defined in Equation~\ref{eq:mean_discon_spec}, for all North/South-polarized antennas in the array (center panel). Just as in Figure~\ref{fig:shape_comparison} we show flagged and unflagged antennas, highlighting example auto-correlation waterfalls of good (85; right panel) and bad (89; left panel) antennas, as defined by the modified $z$-score of their distance metric (Equation~\ref{eq:mean_discon_dist}). The discontinuities are often hard to perceive without a very careful inspection of the waterfall.  Once again, these sorts of effects are often more visible in metric spectra and in renormalized waterfalls, as demonstrated in Figure~\ref{fig:dashboard}.}
    \label{fig:discontinuity_comparison}
\end{figure*}

In Figure~\ref{fig:discontinuity_comparison}, we show metric spectra for all antennas for a single polarization and examples of nominal and abnormal waterfalls. 
Antennas flagged as bad show a wide variety of strange behavior: some show broadband effects, others are more localized. Antenna 89 and one other even shows spectrally oscillatory levels of temporal discontinuities; we currently have no explanation for this effect. Perhaps these features provide further clues to the ongoing system integration and debugging efforts. 

The good antennas are fairly tightly clustered around the average, which is spectrally flat. That behavior is expected if the integration-to-integration differences are purely attributable to thermal noise. Normalizing each waterfall by the average good waterfall should cancel out the spectral and temporal dependence of the noise.  Given that theoretical expectation this might be the easiest of all the metrics to set an absolute cut, rather than a relative one based on the modified $z$-score. However, the wide variety of poorly-understood malfunctions combined with the possibility that low-level RFI might still contaminate these metrics complicates that picture.

\subsection{Assessing Individual Antenna Quality in Practice}\label{sec:practice}

One advantage of the auto-correlation metrics framework is that it is straightforwardly applicable to new combinations of metric spectra and distance measures. For example, it should be noted that the anomalous temporal structure metrics in Sections~\ref{sec:temp_var} and \ref{sec:temp_discon} are not exhaustive. By averaging over the whole night, they privilege frequent or persistent effects over infrequent ones. For example, a strong jump in the waterfall like we see in Antenna 110 in Figure~\ref{fig:variability_comparison} that then quickly reverts to ``standard'' behavior and does not repeat might not be caught by either metric. One could imagine other ways of computing $S(\nu)$ or $D$ that up-weight rare excursions from normality. While we continue to assess antenna malfunctions and develop other metrics, it is  worthwhile to continue the visual inspection of auto-correlation waterfalls normalized by the average of nominally good antennas to identify other modalities of malfunction. 

\begin{figure}[h]
    \centering
    \includegraphics[width=1.0\textwidth]{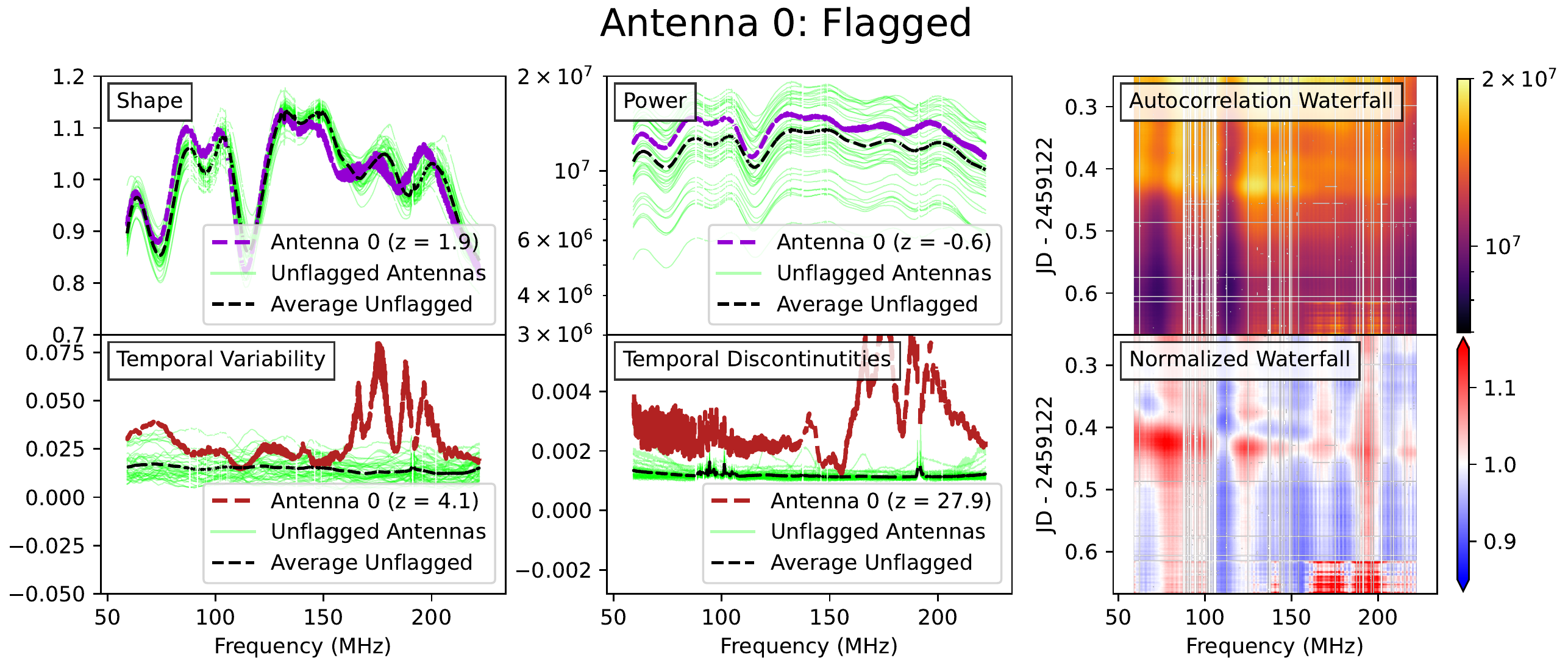}
    \includegraphics[width=1.0\textwidth]{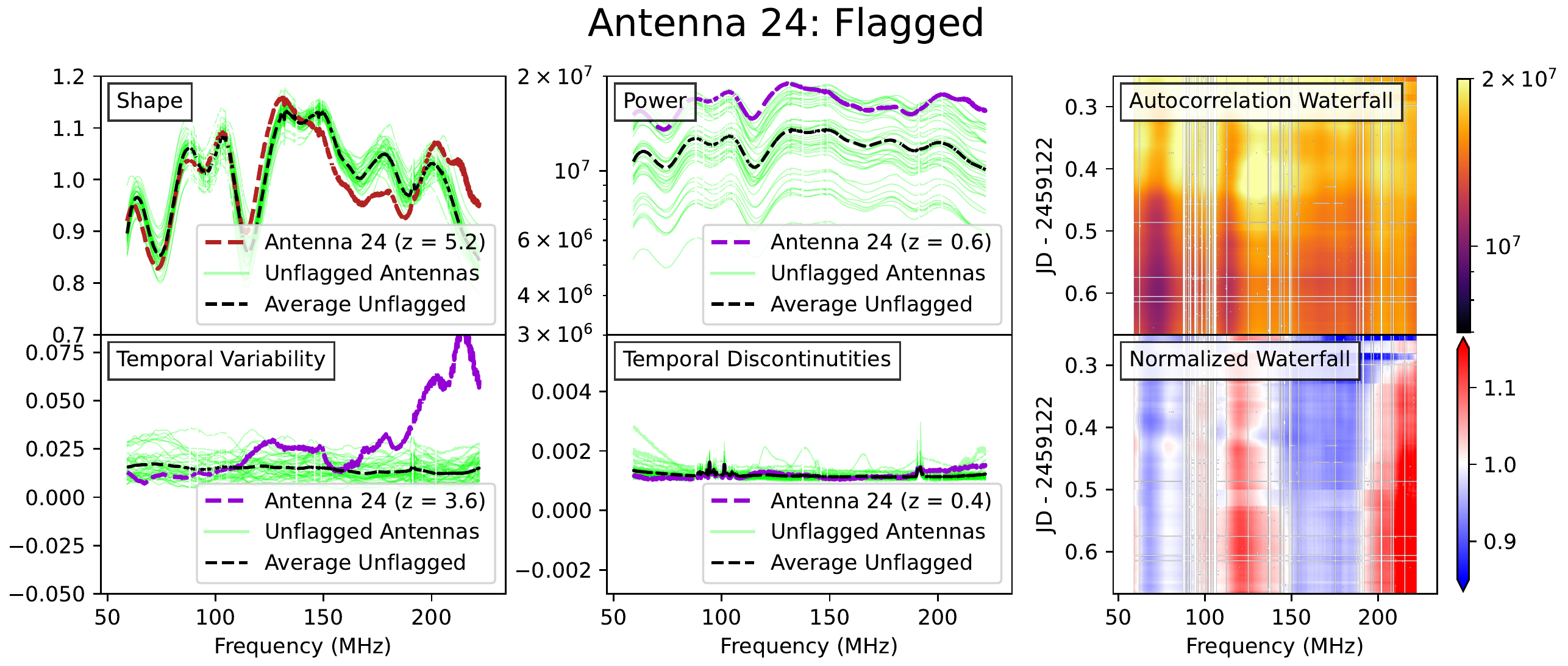}
    \includegraphics[width=1.0\textwidth]{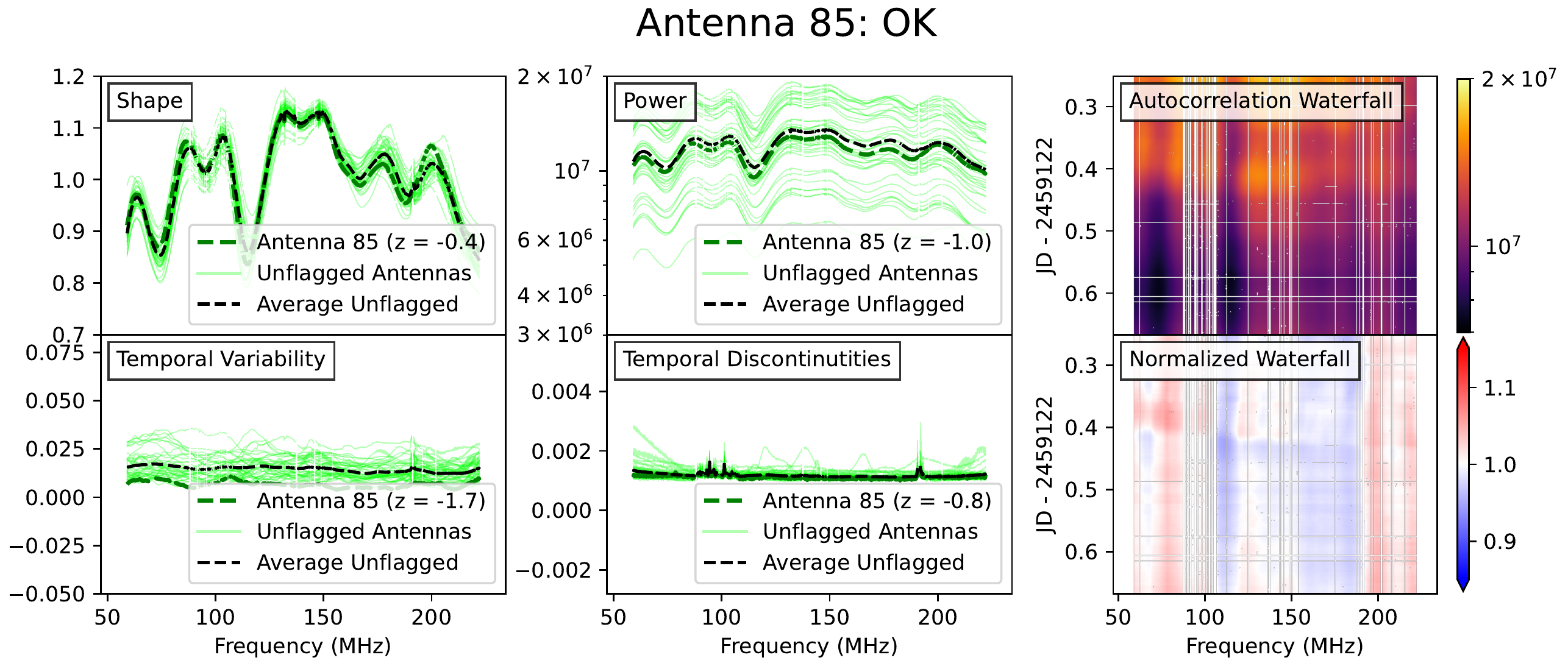}
    \caption{ An example of the summary dashboard used to inspect antenna metrics showing three cases --one for a clearly malfunctioning antenna, one for a borderline flagged antenna, and one for a good antenna. In each we show the metric spectra of the individual antenna compared to all good antennas in light green, helping us to easily see whether the antenna is an outlier. We also show the full auto-correlation waterfalls, both raw and fractional deviation from the antenna average (Figure~\ref{fig:avg_auto_flagged}). The effects detected by our metrics can generally be seen in either the raw or normalized waterfall.
    }
    \label{fig:dashboard}
\end{figure}

In particular, we find it useful to produce a suite of per-antenna visualizations of the different metric spectra and the corresponding auto-correlation waterfalls. In Figure~\ref{fig:dashboard} we show three such examples: one clearly malfunctioning (Antenna 0), one nominal (Antenna 85), and one borderline case that we ultimately flagged (Antenna 24). For each, we show their metric spectra compared to all unflagged antennas, along with the $z$-scores, highlighting which antennas were automatically flagged. These plots synthesize the information about how discrepant each antenna is along the four axes considered here and help clarify why.

In Figure~\ref{fig:dashboard} we also show both the waterfalls and the normalized waterfalls, which are divided by the average good waterfall (Figure~\ref{fig:avg_auto_flagged}) and then normalized to average to 1. We find it particularly useful to look closely at these normalized waterfalls, especially in borderline cases like Antenna 24. Antenna 24's bandpass shape is sufficiently discrepant with the others to merit an automatic flag, though this does not necessarily mean that it is uncalibratable. More concerning are the abrupt discontinuities at high frequency around 2459122.3 and around 2459122.5. This is precisely the kind of issue we worried about: a strong but rare temporal feature that just barely misses the threshold. Examples like this motivate by-eye inspection of borderline antennas. This is what we have done with recent HERA data. The automatic pipeline produces \texttt{jupyter} notebooks with plots like Figure~\ref{fig:dashboard} for all antennas, sorting them by the single highest $z$-score metric. This makes it easy to find the borderline antennas and decide whether to flag them on a case-by-case basis.

\section{Summary}
\label{ch:Summary}

There are a number of current and upcoming interferometers with hundreds of antennas aiming to reach the extreme dynamic range necessary to detect and characterize the neutral hydrogen signal from the epoch of reionization. Separating that signal from foregrounds four to five orders of magnitude stronger requires both large volumes of data and the swift and reliable identification of malfunctions that adversely affect data quality. In this work, we report on new metrics which sensitively detect various pathologies 
and reliably classify them, using HERA data as a case study. In some cases, the precise underlying mechanism (e.g. an antenna with swapped cables for its two polarizations) is known. In others, a physical explanation requires lab and field tests that are beyond the scope of this paper. Armed with per-antenna classifications, instrument teams can more effectively triage issues according to their occurrence rate. In HERA's case, by inspecting the nightly analysis and dashboard reports that implement the metrics outlined here the team can quickly assess the impact of hardware changes. Meanwhile, the definition of metric \emph{spectra} provides a physically meaningful signature which can be exploited by instrument engineers to identify characteristics like reflections, clipping, interference, and more.

The definition of metrics which isolate features of interest and standard ways of displaying them routinely is crucial to managing a large array with a small team. As digital and analog systems grow in capability, arrays will continue to grow in antenna count. Arrays like OVRO-LWA-III \cite{OVRO}, DSA-2000 \cite{dsa2000-aas2021}, HIRAX \cite{HIRAX}, CHORD \cite{chord2019}, PUMA \cite{puma2020}, SKA-Low \cite{SKA} and more will use hundreds to thousands of elements. Ultimately the maintenance time per-antenna imposes a significant design pressure on large arrays. This kind of pressure can also affect arrays with fewer antennas but with more elaborate receivers or wider geographic distributions. A prime example of this regime is the proposed ngVLA \cite{ngVLA}. With 244 antennas distributed across New Mexico, Arizona, and Mexico, along with outriggers extending to VLBA sites across north America and six cryogenic receivers, operation will require careful minimization of maintenance time.\footnote{See ngVLA memo 020.10.05.00.00-0002-PLA, S6.2 at \url{https://ngvla.nrao.edu/page/projdoc}}. Quick identification of subtle systematic errors using semi-automatic systems like we describe here are expected to be essential.

In 21\,cm cosmology experiments, the reliability and precision of arrays will continue to be the dominant factor affecting sensitivity. Identifying, flagging, and ultimately fixing  subtle instrument issues will continue to be the first line of defense. Further work in this area is needed, for example, using simulations to replace the detection of relative outliers with absolute thresholds or to replace iterative flagging with a single analysis step. That said, a system like the one presented here will be necessary for triaging malfunctions and extracting science-quality data to form the basis for future cosmology results.


\acknowledgments
This material is based upon work supported by the National Science Foundation under Grant Nos. 1636646 and 1836019 and institutional support from the HERA collaboration partners. This research is funded by the Gordon and Betty Moore Foundation through grant GBMF5215 to the Massachusetts Institute of Technology. HERA is hosted by the South African Radio Astronomy Observatory, which is a facility of the National Research Foundation, an agency of the Department of Science and Innovation. JSD gratefully acknowledges the support of the NSF AAPF award \#1701536. A. Liu acknowledges support from the New Frontiers in Research Fund Exploration grant program, the Canadian Institute for Advanced Research (CIFAR) Azrieli Global Scholars program, a Natural Sciences and Engineering Research Council of Canada (NSERC) Discovery Grant and a Discovery Launch Supplement, the Sloan Research Fellowship, and the William Dawson Scholarship at McGill. D. Storer acknowledges that this material is based upon work supported by the National Science Foundation Graduate Research Fellowship Program under Grant No. DGE-1762114. Any opinions, findings, and conclusions or recommendations expressed in this material are those of the author(s) and do not necessarily reflect the views of the National Science Foundation. MGS and PK acknowledge support from the South African Radio Astronomy Observatory (SARAO) and National Research Foundation (Grant No. 84156). G.B. acknowledges support from the Ministero degli Affari Esteri della Cooperazione  Internazionale - Direzione Generale per la Promozione del Sistema Paese Progetto di Grande Rilevanza ZA18GR02 and the National Research Foundation of South Africa (Grant Number 113121) as part of the ISARP RADIOSKY2020 Joint Research Scheme.\\ 
\hfill \break
\noindent \textbf{Data Availability} \\
\noindent Data used in this paper is available upon request to the corresponding author. Software used in this paper can be found at \url{https://github.com/HERA-Team/hera_qm}.


%
%

\bibliography{biblio.bib}

%
%
%
%
%

\appendix

\section{Algorithms}
\label{Appendix:algs}

\begin{figure}
    \centering
    \includegraphics[width=\textwidth]{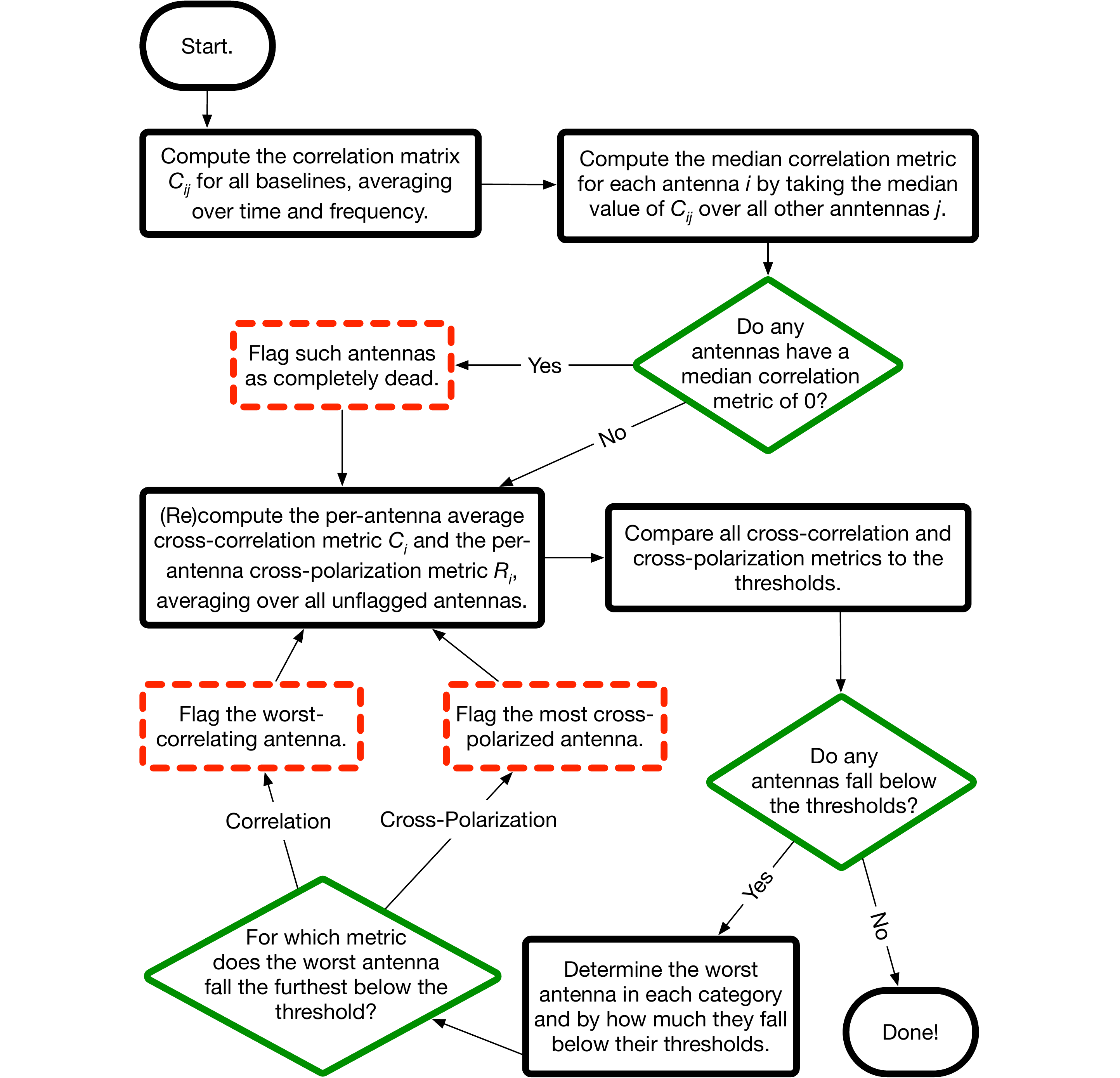}
    \caption{Pseudocode flowchart of the cross-correlation metrics flagging algorithm, as discussed in Section \ref{ch:ant_metrics}.}
    \label{alg:ant_metrics}
\end{figure}

\begin{figure}
    \centering
    \includegraphics[width=\textwidth]{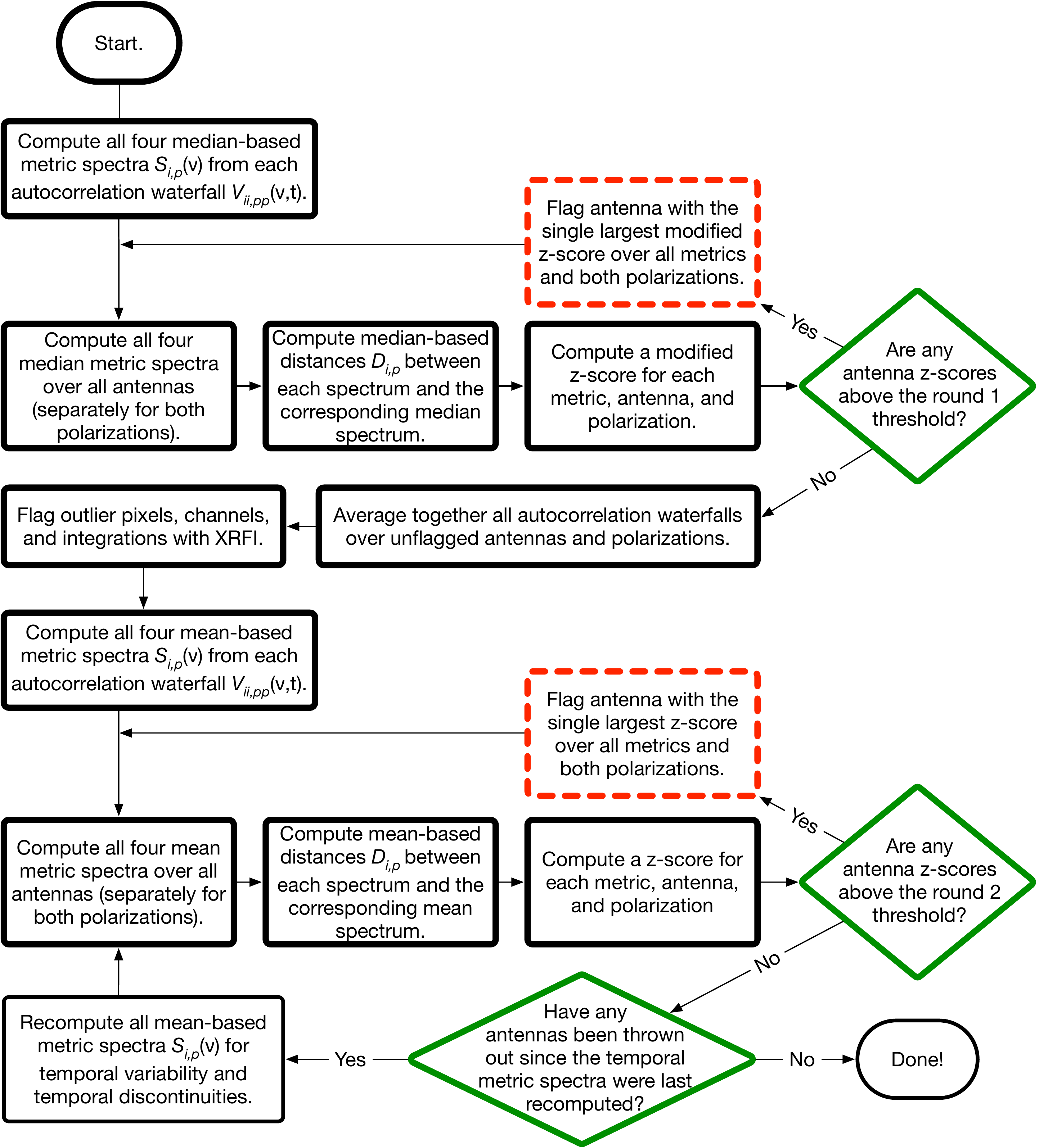}
    \caption{Pseudocode flowchart of the auto-correlation metrics flagging algorithm, as discussed in Section \ref{ch:auto_metrics}.}
    \label{alg:auto_metrics}
\end{figure}

Figure \ref{alg:ant_metrics} provides a pseudocode flowchart for the iterative antenna flagging algorithm described in Section \ref{sec:flagging}. Figure \ref{alg:auto_metrics} provides a  pseudocode flowchart for the auto-correlation based flagging algorithm described in Section \ref{ch:auto_metrics}.

\end{document}